\documentclass[a4paper, 12pt]{article}
\usepackage[utf8]{inputenc}
\usepackage{authblk}
\usepackage{setspace}
\usepackage{graphicx}
\usepackage{float}
\usepackage{listings}
\usepackage{booktabs}
\usepackage{siunitx}
\usepackage{amsmath,amssymb}
\usepackage{fancyvrb}
\usepackage[square, numbers]{natbib}

\usepackage{caption}

\DeclareCaptionFont{mysize}{\fontsize{9}{11}\selectfont}
\usepackage{color}
\usepackage[normalem]{ulem}

\definecolor{ultramarine}{RGB}{0,32,96}

\renewcommand\rightmark{[Agreement tests for 2x2 tables]}

\usepackage{fancyvrb}
\usepackage{varioref,nameref,etoolbox}
\usepackage{hyperref}

\makeatletter 
\newcommand\mynobreakpar{\par\nobreak\@afterheading\vspace{0.1in}} 
\makeatother

\definecolor{britishracinggreen}{rgb}{0.0, 0.26, 0.15}
\hypersetup{
	pdftitle={Agreement coefficients in 2x2 tables},
	pdfauthor={Siqueira JO, Silveira PSP},
	pdfsubject={Statistical Methods},
	pdfkeywords={Cohen's Kappa, Agreement coefficients, Contingency},
	pdfpagemode=UseOutlines,
	bookmarks=true,
	bookmarksopenlevel=3,
    colorlinks=true,
    linkcolor=britishracinggreen,
    citecolor=britishracinggreen,
    urlcolor=britishracinggreen,
}

\title{Better to be in agreement than in bad company: a critical analysis of many kappa-like tests assessing one-million 2x2 contingency tables}
\date{April 12, 2022}

\author[1,2,*]{Paulo Sergio Panse Silveira}
\author[2]{Jose de Oliveira Siqueira}
\affil[1]{Department of Pathology (LIM01-HCFMUSP)}
\affil[2]{Department of Legal Medicine, Medical Ethics, Work and Social Medicine}
\affil[ ]{University of Sao Paulo, SP, Brazil}
\affil[*]{siqueira@usp.br}

\begin{document}

\maketitle 
    
\clearpage
\section*{Abstract}

We assessed several agreement coefficients applied in 2x2~contingency tables, which are commonly applied in research due to dicotomization by the conditions of the subjects (e.g., male or female) or by conveniency of the classification (e.g., traditional thresholds leading to separations in healthy or diseased, exposed or non-exposed, etc.). More extreme table configurations (e.g., high agreement between raters) are also usual, but some of the coefficients have problems with imbalanced tables. Here, we not only studied some especific estimators, but also developed a general method to the study for any estimator candidate to be an agreement measurement. This method was developed in open source R codes and it is avaliable to the researchers. Here, we tested this method by verifying the performance of several traditional estimators over all 1,028,789~tables with size ranging from 1 to 68. Cohen's~kappa showed handicapped behavior similar to Pearson's~r, Yule's~Q, and Yule's~Y. Scott's~pi has ambiguity to assess situations of agreement between raters. Shankar and Bangdiwala's~B was mistaken in all situations of neutrality and when there is greater disagreement between raters. Dice's~F1 and McNemar's~chi-squared incompletely assess the information of the contingency table, showing the poorest performance among all. We concluded that Holley and Guilford's~G is the best agreement estimator, closely followed by Gwet's~AC1 and they should be considered as the first choices for agreement measurement in contingency 2x2~tables. All procedures and data were implemented in R and are available to download from https://sourceforge.net/projects/tables2x2.

\subsection*{keywords}

agreement coefficient\newline
contingency table\newline
categorical data analysis\newline
inter-rater reliability

\subsection*{Current submission}

This manuscript is under consideration of \href{https://www.springer.com/journal/13428}{Behavior Research Methods} since 12Apr2022

\section{Introduction}\label{sec:introduction}

Contingency tables are prevalent and, among them, 2x2 tables reign.
Conclusions drawn from them depend on statistical measures. In Psychology, 
many observations are somewhat fuzzy and, thus, careful
researchers must verify whether different assessors agree, which is
usually an application of Cohen's~$\kappa$ test~\cite{Banerjee1999,Gwet2008,Gwet2010}. In Healthcare, clinical trials
depend on application of drugs, devices or procedures for checking their
association with potential positive or negative effects, which makes Fisher's exact and
chi-squared tests largely used~\cite{Ludbrook2011}. In Epidemiology researchers also need to assess 
agreement between observers, for what they apply McNemar's~$\chi^2$ test~\cite{Kirkwood2003}. Other situations,
concerned with quantifications of effect intensity of potential harmful or
protective expositions to the increased or decreasing existence of diseases
at the populacional level, odds~ratio estimations are traditionally applied~\cite{King2012}.

In all these frequent situations, dicotomization is handy. Sometimes it
is imposed by the study design when a researcher is compared to
another one or when patients are classified in two levels (e.g., male or female).
In other cases, some longed-evaluated cutoff is determined and applied to
label two groups as healthy or diseased, exposed or non-exposed, showing
a given effect or not. It is not a defect of scientific research, but
the recommended most parsimonious approach to start an investigation
with the simplest classification of events. Statistics has the role to
refine researcher impressions, providing evidence pro or against
initial hypotheses. Therefore, it must be useful if statistical
procedures are reliable in a number of situations.

Here were assessed the behavior of several widespread tests 
by exhaustive creation of all possible 2x2
contingency tables for a given range of total sizes (from 1 to 68, 
total of 1,028,789 possible tables) in order to verify
the statistical distribution of their respective measures. It was found
that some estimators may have problems with imbalanced tables due to higher countings
concentrated in one or more table cells.

Before one alleges that this
study is approaching extreme, atypical or unrealistic scenarios, let us
consider that imbalanced tables are the main goal for researchers.
Psychologists desire the highest agreement between assessors while
Clinicians expect the strongest association between intervention and
patient outcome, thus resulting in 2x2 tables with data concentration
along the main or off-diagonal. Epidemiologists, likewise, pursue
situations of high association between exposition and effect on populational 
diseases, usually low in prevalence, results in 2x2 tables with relatively 
empty cells corresponding to affected people.

Conversely, balanced tables improve little scientific knowledge for
they would show, respectively, inability to obtain consistent measures
from a given method, absence of treatment effect or lack of relationship
between exposition and effect. No one starts a scientific study unless
the hypothesis is solid enough and the method is designed to control
confusion variables. When, after all the best researcher efforts, the
2x2 table results well-balanced, statistical tests will show
non-rejection of null hypothesis and the study failed to confirm one's
expectation, remaining dubious conclusions either by lack of effect or
sample insufficiency.

These few examples show that imbalanced tables are rule, not the exception. 
Estimators for agreement or disagreement between observations or observers 
(generically named as `raters' along this text) must be robust, therefore, 
to extreme tables in order to provide inferential statistics support to the
researchers. 

Here we exhaustively analysed all possible tables with total size from 1 to 68, totalling 1,028,789~different contingency tables to create a comprehensive behavior map of several traditional association or agreement coefficients: Cohen's~$\kappa$, Holley and Guilford's~$G$, Yule's~$Q$, Yules’s~$Y$, Pearson's~$r$, McNemar's~$\chi^2$, Scott's~$\pi$, Dice's~$F1$, Shankar and Bangdiwala's~$B$, and Gwet's~$AC1$ in order to show that $G$ and $AC1$ are the best agreement coefficients while the traditional $\kappa$ and McNemar's~$\chi^2$ are problematic.

\section{Methods}\label{sec:methods}

All simulations and essential estimators presented here were implemented in R, 
a free statistical language, and provided to allow replication of our findings.

To simplify mathematical notation along this text, a convention was
adopted according to Table~\ref{tab:mathnotation}.

\begin{table}[H]
\centering\begingroup\fontsize{10}{10}\selectfont
\caption{$A$ and $B$ are representations of events
(positive observer evaluation, existence of disease, exposition or
effect, etc.) and $\bar{A}$ and $\bar{B}$ are their respective
negations (negative observation, health individuals, absence of
exposition or effect, etc.). In the main diagonal, $a$ and $d$ are
counting or proportions of positive and negative agreements. In the
off-diagonal, $b$ and $c$ are counting or
proportions of disagreements. Sample size along this text is $n=a+b+c+d$}
\label{tab:mathnotation}
\begin{tabular}{cccc}
\toprule
 & $B$ & $\bar{B}$ & \\
\midrule
$A$ & $a$ & $b$ & $a+b$ \\
$\bar{A}$ & $c$ & $d$ & $c+d$ \\
\midrule
 & $a+c$ & $b+d$ & $a+b+c+d$ \\
\bottomrule
\end{tabular}
\endgroup{}
\end{table}

\subsection{Cohen's kappa}\label{sec:cohens-kappa}

This test was originally published by Cohen in 1960~\cite{Cohen1960}. The original
publication proposed a statistical method to verify agreement, typically
applied to comparison of observers or results of two measurement methods
such as laboratory results (raters). Kappa ($\kappa$) statistics is given by
\begin{eqnarray}
\label{eq:kappa}
\begin{aligned}[b]
\kappa = \frac{p_o - p_c}{1 - p_c}
\end{aligned}
\end{eqnarray}
where \(p_o$ is the realized proportion of agreement and $p_c$ is the expected proportions by chance (i.e., under assumption of null hypothesis) in which both raters agreed (i.e., the sum of proportions along the matrix main diagonal). 

For 2x2 tables, in more concrete terms, 
\begin{eqnarray}
\label{eq:popc}
\begin{aligned}[b]
& p_o={\frac{a+d}{n}} &\\
& p_c={\frac{(a+b)(a+c)+(c+d)(b+d)}{n^2}} &
\end{aligned}
\end{eqnarray}

Thus, Cohen's $\kappa$ can be computed by

\begin{eqnarray}
\label{eq:kappa_abcd}
\begin{aligned}[b]
\kappa = {\frac{2(ad-bc)}{(a+c)(c+d)+(b+d)(a+b)}}
\end{aligned}
\end{eqnarray}

This equation shows tension between the main ($ad$) and off ($bc$) diagonals. The greater the agreement (or the smaller the disagreement) the greater the value of $\kappa$.

In addition, Cohen showed that kappa statistics has a maximal value
permitted by the marginals:
\begin{eqnarray}
\label{eq:kappamax}
\begin{aligned}[b]
\kappa_{M} = { \frac{p_{oM} - p_c}{1 - p_c} }
\end{aligned}
\end{eqnarray}

where $p_{oM}$ is the sum of minimal marginal values taken in pairs. For 2x2 tables it is 
\begin{eqnarray}
\label{eq:pomax}
\begin{aligned}[b]
p_{oM} = { min\left( (a+c),(a+b) \right) + min\left( (b+d),(c+d) \right) }\end{aligned}
\end{eqnarray}

When the intensity of agreement is to be qualified~(Table~\ref{tab:kappacategory}), Cohen's recommendation is to compute $\kappa/\kappa_M$ to correct the agreement value of $\kappa$, but it seems to be largely forgotten by researchers~\cite{Sim2005}. An additional complication is that criteria varies according to different authors~\cite{Wongpakaran2013}. Cohen did not propose correction by any lower bound when $\kappa<0$~(i.e., rater disagreement), stating that it is more complicated and depends on the marginal values. For that reason, in case of disagreement no correction was performed in this work.

\begin{table}[H]

\caption{criteria for qualitative categorization of $\kappa$ according different authors~---~modified from Wongpakaran et al., 2013~\cite{Wongpakaran2013}.}
\label{tab:kappacategory}
\centering
\resizebox{\linewidth}{!}{
\fontsize{7}{9}\selectfont
\begin{tabular}[t]{cc|cc|cc}
\toprule
$\kappa$ & Landis and Koch & $\kappa$ & Altman & $\kappa$ & Fleiss\\
\midrule
{}[-1.0,0.0) & Poor &   &   &   &  \\
{}[0.0,0.2) & Slight & {}[-1.0,0.2) & Poor &   &  \\
{}[0.2,0.4) & Fair & {}[0.2,0.4) & Fair & {}[-1.0,0.4) & Poor\\
{}[0.4,0.6) & Moderate & {}[0.4,0.6) & Moderate & \\
{}[0.6,0.8) & Substantial & {}[0.6,0.8) & Good & {}[0.4,0.75) & Intermediate to Good \\
{}[0.8,1.0] & Almost perfect & {}[0.8,1.0] & Very good & {}[0.75,1.0] & Excellent\\
\bottomrule
\end{tabular}}
\end{table}

\subsection{Concurrent agreement measures}\label{sec:concurrent}

Many other alternative statistics have been proposed to compute agreement,
although not always originally conceived for this purpose. Besides Cohen's~$\kappa$, here
we selected Holley and Guilford's~$G$, Yule's~$Q$, Yules's~$Y$,
Pearson's~$r$, McNemar's~$\chi^2$, Scott's $\pi$, Dice's $F1$, 
Shankar and Bangdiwala's~$B$, and Gwet's $AC1$.

Some other estimators are redundant and were not analyzed for varied reasons:\mynobreakpar
\begin{itemize}
\item Janson and Vangelius'~$J$, 
Daniel-Kendall’s generalized correlation coefficient, 
Vegelius’ E-correlation, and Hubert's~$\Gamma$~(see~``\nameref{tit:G}");
\item Goodman-Kruskal's~$\gamma$, odds-ratio, and risk-ratio~(see~``\nameref{tit:Q}");
\item Pearson's~$\chi^2$, Yule's~$\phi$, Cramér's~$V$, Matthews' correlation 
coefficient, and Pearson's contingent coefficient~(see~``\nameref{tit:V}"~and~``\nameref{tit:r}");
\item Fleiss'~$\kappa$~(see section~``\nameref{tit:scott_pi}").
\end{itemize}

All these alternatives are effect-size measures, therefore independent
of sample size, $n$. A brief description of each test in the present context follows.

\subsubsection{Holley and Guilford's $G$}\label{tit:G}

One of the simplest approach to a 2x2~table was proposed by Holley \& Guilford, 1964~\cite{Holley1964}, given by\mynobreakpar

\begin{eqnarray}
\label{eq:G}
\begin{aligned}[b]
G = \frac{(a+d)-(b+c)}{a+b+c+d}
\end{aligned}
\end{eqnarray}

A generalized agreement index is $J$~\cite{Janson1982} that can be applied to larger tables. However, in 2x2~tables it reduces to $J = G^2$, thus $J$ performance was excluded from the current analysis. 

Other proposed coefficiente, Hubert's $\Gamma$~\cite{Hubert1977} is a special case of Daniel-Kendall's generalized correlation coefficient and Vegelius' E-correlation~\cite{Janson1982}. For 2x2~tables it is computed by\mynobreakpar
\begin{eqnarray}
\label{eq:hubert_gamma}
\begin{aligned}[b]
\Gamma = { 1 - 4 \frac{(a+d)(b+c)}{n^2}}
\end{aligned}
\end{eqnarray}

Although it looks like another coeffient, it is possible to show its equivalency to:
\mynobreakpar
\begin{eqnarray}
\label{eq:hubert_gamma2}
\begin{aligned}[b]
\Gamma = {\left( {\frac{ (a+d)-(b+c)}{a+b+c+d}} \right)^2 } = G^2
\end{aligned}
\end{eqnarray}

Since it is redundant to both $J$ and Holley and Guilford $G$, the analysis of $\Gamma$ coefficient is also not required.

\subsubsection{Yule's Q}\label{tit:Q}

Goodman-Kruskal's $\gamma$ measures association between ordinal variables. In the special case of 2x2 tables Goodman-Kruskal's~$\gamma$ corresponds to Yule's~Q~\cite{Yule1912}, also known as Yule's coefficient of association. It can be computed by\mynobreakpar 

\begin{eqnarray}
\label{eq:Q}
\begin{aligned}[b]
Q = {{ad-bc}\over{ad+bc}}
\end{aligned}
\end{eqnarray}

Yule's~$Q$ is also related to odds~ratio, which was not conceived nor applied as an
agreement measure (although it could be). The relationship is\mynobreakpar

\begin{eqnarray}
\label{eq:OR}
\begin{aligned}[b]
OR = {{ad}\over{bc}} =  {{1+Q}\over{1-Q}}
\end{aligned}
\end{eqnarray}

It is recommended to express $OR$ as logarithm. Therefore,\mynobreakpar

\begin{eqnarray}
\label{eq:logOR}
\begin{aligned}[b]
log(OR) = log(ad)-log(bc) 
\end{aligned}
\end{eqnarray}

Again, it is possible to observe that $OR$ and $Q$ are statistics from the same tension-between-diagonals family.

Risk ratio ($RR$) also belongs to this family, being defined (assuming
exposition in rows and outcome in columns of a 2x2~table) as the probability of outcome among exposed individuals relative to the probability of outcome among non-exposed individuals. Since $RR$ describes only probability ratio of occurence of outcomes, we propose to define it as a positive risk ratio, computed by\mynobreakpar
\begin{eqnarray}
\label{eq:RRpositive}
\begin{aligned}[b]
RR_{+} = \frac{\frac{a}{a+b}}{\frac{c}{c+d}}
\end{aligned}
\end{eqnarray}

To our knowledge, it is not usual in epidemiology the definition of a
negative risk ratio (the ratio between the probabilities of absence of outcome among exposed individuals and absence of outcome among non-exposed individuals), which should be conceived as\mynobreakpar
\begin{eqnarray}
\label{eq:RRnegative}
\begin{aligned}[b]
RR_{-} = \frac{\frac{b}{a+b}}{\frac{d}{c+d}}
\end{aligned}
\end{eqnarray}

Consequently:
\begin{eqnarray}
\label{eq:ORasRRs}
\begin{aligned}[b]
OR = {{RR_{+}} \over {RR_{-}} } = { {ad} \over {bc} }
\end{aligned}
\end{eqnarray}

Therefore, it is arguable that the traditional $RR_{+}$ is a somewhat incomplete measure of agreement, for it does not explore all the information of a 2x2~table when compared to $OR$.

Both $RR$ and $OR$ are transformations of $Q$, inheriting their characteristics. For that reason, only $Q$ is analyzed in this work.

\subsubsection{Yules's~$Y$}

The coefficient of colligation, $Y$, was also developed by~Yule, 1912~\cite{Yule1912}.
It is computed by\mynobreakpar
\begin{eqnarray}
\label{eq:Y}
\begin{aligned}[b]
Y = \frac{\sqrt{ad}-\sqrt{bc}}{\sqrt{ad}+\sqrt{bc}}
\end{aligned}
\end{eqnarray}
which is a variant of Yule's~$Q$. Here, each term can be interpreted as a geometric mean.

\subsubsection{Cramér's V}\label{tit:V}

The traditional Pearson's chi-squared~($\chi^2$) test can be computed by
\begin{eqnarray}
\label{eq:X2}
\begin{aligned}[b]
X^2={ {(ad-bc)^2(a+b+c+d)}\over{(a+b)(c+d)(a+c)(b+d)} }
\end{aligned}
\end{eqnarray}

This formulation is interesting to reveal $\chi^2$~statistics containing
tension between diagonals, $ad-bc$. 

For the special case of 2x2 tables, absolute value of $\kappa$ and $\chi^2$ are associated~\cite{Feingold1992}. However, it is observed that $\chi^2$~statistics is not an effect size measurement because it depends on sample sizes thus, in its pure form, $\chi^2$ does not belong to the agreement-family of coefficients. In order to remove sample size dependence and turn $\chi^2$~statistics into an effect size measurement, it should be divided by $n = a+b+c+d$. The squared root of this transformation is Cramér's $V$~\cite{Cramer1946}, computed by\mynobreakpar
\begin{eqnarray}
\label{eq:V}
\begin{aligned}[b]
V = {\sqrt{X^2 \over n}} = { {|ad-bc|} \over {\sqrt{(a+b)(c+d)(a+c)(b+d)}}  }
\end{aligned}
\end{eqnarray}

Cramér's~$V$ can also be regarded as the absolute value of an implicit Pearson's correlation between nominal variables or, in other words, an effect size measure ranging from 0 to 1. Cramer's~$V$, in other words, is the tension between diagonals, $ad-bc$ (which is the 2x2~matrix determinant) normalized by the productory of marginals $(a+b)(c+d)(a+c)(b+d)$. 

\subsubsection{Pearson's $r$}\label{tit:r}

There are many relations and mathematical identities among coefficients that converge to Pearson's correlation coefficient,~$r$. 

Matthews' correlation coefficient is a measure of association between dichotomous variables~\cite{Matthews1975} also based on $\chi^2$~statistics. Since it is defined from the assessment of true and false positives and negatives, it is regarded as a measure of agreement between measurement methods. It happens that Matthews' correlation coefficient is identical to Pearson's~$\phi$ coefficient and Yule's~$\phi$ coefficient~\cite{Cohen1968}, computed by\mynobreakpar
\begin{eqnarray}
\label{eq:phi}
\begin{aligned}[b]
\phi = {{ad-bc} \over \sqrt{(a+b)(a+c)(b+d)(c+d)}}
\end{aligned}
\end{eqnarray}

Another famous estimator is the Pearson's contingent coefficient, usually defined from chi-squared statistics and also expressed as function of $\phi$ by\mynobreakpar 
\begin{eqnarray}
\label{eq:PCC}
\begin{aligned}[b]
PCC = \sqrt{\frac{X^2}{X^2+n}} = \sqrt{\frac{\phi^2}{\phi^2+1}}
\end{aligned}
\end{eqnarray}
It was not included in this analysis for two reasons: it is not taken as an agreement coefficient and its value ranges from zero to $\sqrt{\frac{1}{2}}\approx0.707$, which makes this coefficient not promissing to the current context.

Other two correlation coefficients, Spearman's $\rho$ and Kendall's $\tau$, also provide the same values of Pearson's~$r$ for 2x2 tables. In the notation adopted here:
\begin{eqnarray}
\label{eq:r}
\begin{aligned}[b]
r = \rho = \tau = { {ad-bc} \over { \sqrt{(a+b)(a+c)(b+d)(c+d)} } }
\end{aligned}
\end{eqnarray}

From equation~\ref{eq:r} other coincidences are observed:\mynobreakpar
\begin{itemize}
\item equation~\ref{eq:phi} shows that $\phi=r$ for 2x2 tables, thus Matthews' correlation coefficient and Pearson's contingent coefficient also share $r$~properties.
\item equation~\ref{eq:V} shows that Cramér's~$V$ merely is the absolute value of Pearson's $\phi$ coefficient~\cite{Matthews1975}, which is, in turn, equal~to~$r$. 
\end{itemize} 

Consequently, for the current work only Pearson's~$r$ is computed as representative of all these other estimators. 

\subsubsection{McNemar's chi-squared}\label{mcnemars-chi-squared}

This test was created by McNemar, 1947~\cite{McNemar1947}, became known in the literature as McNemar's~$\chi^2$. It is applicable to 2x2 tables by
\begin{eqnarray}
\label{eq:MNX2}
\begin{aligned}[b]
X^2_{MN}={{(b-c)^2}\over{b+c}}
\end{aligned}
\end{eqnarray}
to assess marginal homogeneity. typical example is the verification of change before and after an intervention, such as the disappearance of a disease under treatment in
a given number of subjects. 

The traditional McNemar's~$\chi^2$ cannot be directly confronted with other estimators because it is not restricted to the interval [-1,1]. It can be normalized to show values between 0 and 1 as an effect-size measurement if divided by $|b-c|$, resulting in \mynobreakpar
\begin{eqnarray}
\label{eq:MN}
\begin{aligned}[b]
MN={{{(b-c)^2}\over{b+c}}\over{|b-c|}}={{|b-c|}\over{b+c}}
\end{aligned}
\end{eqnarray}

It is noteworthy to say that both McNemar's~$\chi^2$ and its normalized correspondent $MN$ are even more partial than $RR$, for they use only the information of the off-diagonal. Problems caused by such a weakness are explored below. 

\subsubsection{Scott's pi}\label{tit:scott_pi}

This statistics is similar to Cohen's~$\kappa$ to measure inter-rater reliability
for nominal variables~\cite{Scott1955}. It applies the same equation of $\kappa$ but it changes the estimation of $p_c$ using squared joint proportions,
computed as 
\begin{eqnarray}
\label{eq:pi_as_k}
\begin{aligned}[b]
\pi = { {p_o - p_c}\over{1 - p_c} }
\end{aligned}
\end{eqnarray}
 
where
\begin{eqnarray}
\label{eq:pi_pc}
\begin{aligned}[b]
p_c = { \left ( {{a+c+a+b} \over {2 n}} \right )^2 + \left ( {{c+d+b+d} \over {2 n}} \right )^2 }
\end{aligned}
\end{eqnarray}

thus
\begin{eqnarray}
\label{eq:pi}
\begin{aligned}[b]
\pi = {  {ad- ( {{b+c} \over 2}}  )^2 \over {({a+ {{b+c} \over 2}}) ({d+ {{b+c} \over 2}}}) }
\end{aligned}
\end{eqnarray}

Besides the tension between the diagonals shown on the numerator of
this expression ($ad$ vs.~$b+c$), Scott's $\pi$ also coincides with
Fleiss' $\kappa$ in the special case of 2x2 tables, thus our analysis
is restricted to Scott's~$\pi$.

\subsubsection{Dice's $F1$}\label{dices-f1}

Also known as F-score or F-measure, it was first developed by Dice, 1945~\cite{Dice1945} as a measure of test accuracy, therefore it can be assumed as an agreement statistics in the same sense of the Matthews'~correlation coefficient described above. 

It is computed by 
\begin{eqnarray}
\label{eq:F1}
\begin{aligned}[b]
F1 = { {a}\over{a+{{b+c}\over2 }} } = \frac{2a}{2a+b+c}
\end{aligned}
\end{eqnarray}

$F1$ has been suggested to be a suitable agreement measure to replace Cohen's $\kappa$ in medical situations~\cite{Hripcsak2005}, thus its analysis was included here.

There are fundamental differences among $F1$ and all other agreement statistics. It does not belong to the tension-between-diagonals family, for it computes only the proportion between positive agreement ($a$) and the upper-left triangle of a 2x2 table.

In addition, $F1$ is difficult to compare \textit{a priori} with other measurements because it ranges from $F1=0$ if $a=0$ (disagreement) to $F1=1$ if $b=0$ and $c=0$ (agreement) in both cases neglecting agreements in negative countings~($d$). 
Being the range of $F1$ shorter than that of other measurements, neutral situations (i.e., when there is no agreement nor disagreement) should find $F1 \approx 0.5$, while the concurrent measurements presented here should provided zero. 
In order to make its range more comparable, we propose a rescalling to the interval $[-1,1]$ given by\mynobreakpar
\begin{eqnarray}
\label{eq:F1adj}
\begin{aligned}[b]
F1_{adj} = { 2F1 - 1} = \frac{2a-(b+c)}{2a+(b+c)}
\end{aligned}
\end{eqnarray}
which adjusts $F1$ range to span from -1~to~1 without changing its original behavior. Interestingly, this rescalling created a partial tension between the positive agreement and the off-diagonal that was not present in $F1$ original presentation.

\subsubsection{Shankar and Bangdiwala's $B$}\label{Bangdiwala_B}

This coefficient was proposed this statistics to access 2x2~tables, reporting its
good behavior~\cite{Shankar2014}. This statistics is computed by\mynobreakpar

\begin{eqnarray}
\label{eq:B}
\begin{aligned}[b]
B = \frac{a^2+d^2}{(a+c)(a+b)+(b+d)(c+d)}
\end{aligned}
\end{eqnarray}

Similarly to Dice's~$F1$, this estimator ranges from 0 to 1, being 0 correspondent to
disagreement, 0.5 to neutrality, and 1 to agreement. Therefore, we also propose to explore
its adjustment by scaling to the range $[-1,1]$ with:\mynobreakpar

\begin{eqnarray}
\label{eq:B2}
\begin{aligned}[b]
B_{adj} = 2B-1 = {{a^2+d^2- \left( 2bc+(a+d)(b+c) \right)}\over{a^2+d^2+ \left( 2bc+(a+d)(b+c) \right) }}
\end{aligned}
\end{eqnarray}

\subsubsection{Gwet's $AC1$}\label{gwets-ac1}

This first-order agreement coefficient ($AC1$) was developed by~Gwet, 2008~\cite{Gwet2008} as an attempt to correct Cohen's~$\kappa$ distortions when there is
high or low agreement. $AC1$ seems to have better performance than
Cohen's $\kappa$ assessing inter-rater reliability analysis of personality
disorders~\cite{Wongpakaran2013,Xie2017} and has been applied in information
retrieval~\cite{Manning2008}.

It is computed by
\begin{eqnarray}
\label{eq:AC1}
\begin{aligned}[b]
AC1 = { {a^2+d^2- {{(b+c)^2}\over{2}} }\over{ a^2 + d^2 + {{(b+c)^2}\over{2}}} + (a+d)(b+c)}
\end{aligned}
\end{eqnarray}

$AC1$ somewhat reflects the tension between diagonals, since there is added
values for $a$ and $d$ and subtracted values of $b$ and $c$ in the numerator.

\subsection{Performance analysis}\label{sec:performance-analysis}

In ``\nameref{sec:challenge_tables}" we develop the comparative analysis of all these statistics first by challenging them with fabricated 2x2 tables, considering scenarios with more balanced tables, and then with more extreme tables containing 0 or 1 in some cells. These analyses confront intuition of agreement and the measures among all estimatives.

A second step is the analysis of ``\nameref{sec:n64}", verifying the coincidence and stability of statistical decisions taken from the estimators.

Finally, in ``\nameref{sec:n1to68}" we exhaustively mapped all estimatives from all possible tables created in this range of sizes to show the entire region that each concurrent estimator covers. We choose to compute all tables up to $n=68$ for it is the mininum number that generates more than one million (exactly 1,028,789) different tables.

\section{Results}\label{sec:results}

\subsection{Challenge tables}\label{sec:challenge_tables}

A set of 17 tables was chosen to represent several situations: eight situations of 
agreement, two of neutrality and seven of disagreement in several 2x2~configurations. 

In Table~\ref{tab:challenge} regular situations were tested. The intention here is to confront one's intuition with values provided by all estimators assessed in the current work.

\begin{table}[H]
\centering
\caption{test performance of estimators with 2x2 challenge tables showing three levels of agreement, three levels of disagreement, two neutral situations, and a parallel situation were $b = 2a$ and $d = 2c$ (boldface showing discrepancies among tests).}
\label{tab:challenge}
\resizebox{\textwidth}{!}{%
\begin{tabular}{lcccccccccc}
\toprule
 &  & agr.high & agr.high & agr.low & dis.high & dis.high & dis.low & neutral & neutral & dis. a/c=b/d \\
\midrule
 & & $\left[ \begin{array}{cc} 90 & 10  \\ 10 & 90 \end{array}\right]$ &  $\left[ \begin{array}{cc} 90 & 11  \\ 9 & 90 \end{array}\right]$ & $\left[ \begin{array}{cc} 60 & 41  \\ 39 & 60 \end{array}\right]$ & {}$\left[ \begin{array}{cc} 10 & 90  \\ 90 & 10 \end{array}\right]$ & $\left[ \begin{array}{cc} 10 & 91  \\ 89 & 10 \end{array}\right]$ & $\left[ \begin{array}{cc} 41 & 60  \\ 60 & 39 \end{array}\right]$ & $\left[ \begin{array}{cc} 50 & 50  \\ 50 & 50 \end{array}\right]$ & $\left[ \begin{array}{cc} 75 & 25  \\ 75 & 25 \end{array}\right]$ & $\left[ \begin{array}{cc} 44 & 88  \\ 22 & 44 \end{array}\right]$ \\
\midrule
 Holley and Guilford's $G$ &	0/8 &	0.80000 &	0.80000 &	0.20000 &	-0.80000 &	-0.80000 &	-0.20000 &	0.00000 &	0.00000 & -0.11111\\
Gwet's $AC1$ &	1/9 &	0.80000 &	0.80000 &	0.20000 &	-0.80000 &	-0.80000 &	-0.19988 &	0.00000 &	\textbf{0.05882} & -0.11111 \\
Scott's $\pi$ &	1/9 &	0.80000 &	0.80000 &	0.20000 &	-0.80000 &	-0.80000 &	-0.20012 &	0.00000 &	\textbf{-0.06667} & -0.11111 \\
Cohen's $\kappa$ &	1/9 &	0.80000 &	0.80002 &	0.20008 &	-0.80000 &	-0.79982 &	-0.20012 &	0.00000 &	0.00000  & \textbf{0.00000} \\
Corrected Cohen's $\kappa$  &	4/9 &	\textbf{1.00000} &	\textbf{0.98000} &	\textbf{0.98000} &	-0.80000 &	-0.79982 &	-0.20012 &	0.00000 &	0.00000 & \textbf{0.00000} \\
Pearson's $r$ &	1/9 &	0.80000 &	0.80018 &	0.20012 &	-0.80000 &	-0.79998 &	-0.20012 &	0.00000 &	0.00000  & \textbf{0.00000} \\
Yule's $Q$ &	7/9 &	\textbf{0.97561} &	\textbf{0.97585} &	\textbf{0.38488} &	\textbf{-0.97561} &	\textbf{-0.97561} &	\textbf{-0.38488} &	0.00000 &	0.00000  & \textbf{0.00000} \\
Yule's $Y$ &	1/9 &	0.80000 &	0.80090 &	0.20015 &	-0.80000 &	-0.79999 &	-0.20015 &	0.00000 &	0.00000  & \textbf{0.00000} \\
Shankar and Bangdiwala's $B$, &	& 0.81000 &	0.81008 &	0.36004 &	0.01000 &	0.01000 &	0.16008 &	0.25000 &	0.31250 & 0.22222\\
verified by adjusted $B$ &	9/9 & \textbf{0.62000} & \textbf{0.62016} & \textbf{-0.27993} & \textbf{-0.98000} & \textbf{-0.98000} & \textbf{-0.67983} & \textbf{-0.50000} & \textbf{-0.37500} & \textbf{-0.55556} \\
Dice's $F1$, &	&	0.90000 &	0.90000 &	0.60000 &	0.10000 &	0.10000 &	0.40594 &	0.50000 &	0.60000 & 0.44444 \\
verified by adjusted $F1$ &	1/9 &	0.80000 &	0.80000 &	0.20000 &	-0.80000 &	-0.80000 &	-0.18812 &	0.00000 &	\textbf{0.20000} & -0.11111 \\
Normalized McNemar's $\chi^2$ &	8/9 &	\textbf{0.00000} &	\textbf{0.10000} &	\textbf{0.02500} &	\textbf{0.00000} &	\textbf{0.01111} &	\textbf{0.00000} &	0.00000 &	\textbf{0.50000} &	\textbf{0.60000} \\
\bottomrule
\end{tabular}%
}
\end{table}

It may interesting to observe table~\ref{tab:challenge} by columns:\mynobreakpar

\begin{itemize}
\item
  This table indicates test names and countings of the number of problems to perform with the selected 9 scenarios.
\item
  Columns 1 and 2 show scenarios of high agreement.
  Many tests provide a reasonable value, except Corrected Cohen's $\kappa$ (as recomended by the original author) and Yule's $Q$ (exagerated values), Shankar and Bangdiwala's $B$ (underestimated when adjust is applied), and Normalized McNemar's $\chi^2$ (estimated as zero when $b=c$ or underestimated when $b$ and $c$ are too close).
\item
  Column 3 shows a low agreement, for what Normalized McNemar's $\chi^2$ underestimated while Corrected Cohen's~$\kappa$ and Yule's $Q$ overestimated values in comparison with other tests. Adjusted Shankar and Bangdiwala's $B$ mistakenly pointed this table as disagreement.
\item
  Columns 4 to 6 provide disagreement situations that are the reverse of the previous three columns. Yule's $Q$ reveals the same exaggeration for high disagreement (Cohen's~$\kappa$ has no proposed correction for negative values). Normalized McNemar's $\chi^2$ and Shankar and Bangdiwala's $B$ provide only positive values. If Normalized McNemar's $\chi^2$ is taken by its absolute number, disagreement was underestimated. Rescalled Shankar and Bangdiwala's $B$ shows that disagreement was overestimated.
\item
  Column 7 has all values equal. Raw Dice's $F1$ is equal to 0.5 (as expected), provinding zero when rescaled. Shankar and Bangdiwala's $B$, however, could not deal with this completly neutral situation, showing disagreement when rescalled.
\item
  Column 8 caused major problems for Normalized McNemar's $\chi^2$, Shankar and Bangdiwala's $B$ and Dice's~$F1$, while Scott's $\pi$ and Gwet's $AC1$ slightly deviated from zero.
\item
  Column 9 is a special situation os low disagreement that caused problems to many estimators. The matrix determinant of a parallel 2x2 table is null and all estimators belonging to the tension-between-diagonal family become, therefore, null. The slight disagreement was equaly captured by $G$, Gwet's $AC1$, Scott's $\pi$ and adjusted $F1$. Adjusted Shankar and Bangdiwala's~$B$ and Normalized McNemar's~$\chi^2$ overestimated the disagreement.
\end{itemize}

Table~\ref{tab:extremes} shows a second set of unbalanced tables, with the presence of values 0 or 1 in some table cells to provide more extreme conditions:\mynobreakpar
\begin{itemize}
\item
  Many estimators are problematic. Cohen's $\kappa$, $\kappa$ adjusted by maximum $\kappa$, Pearson's $r$, Yule's $Q$ and $Y$  present problems when there are zeros in some cells, providing null or non-computable estimatives. $Q$ and $Y$ easily approached 1 or -1 even when the agreement or disagreement are not perfect. Normalized McNemar's $\chi^2$ uses information only from the off-diagonal and it is disturbed when information is concentraded in the main diagonal. Despite its adjustment, Shankar and Bangdiwala's $B$ failed in some situations of disagreement.
\item
  The second major cause of problems are due to 0 in the main diagonal (last four columns), leading to underestimation of agreement by Pearson's $r$, Cohen's $\kappa$ and Scott's $\pi$, and underestimation of disagreement by Pearson's $r$ and Cohen's $\kappa$.
\item
  When 0 appears in both diagonals (last two columns), many estimators are not computable while others produce underestimated values. Scott's $\pi$ underestimates agreement. Normalized McNemar's~$\chi^2$ overestimated agreement and disagreement. Holley and Guilford's~$G$, Gwet's~$AC1$, Dice's~$F1$, and Shankar and Bangdiwala's $B$ were able to generate adequate values in both situations.
\item
  When there are no zeros (first three columns), still Yules'$Q$ and $Y$ may overestimate agreement of disagreement. Adjusted Shankar and Bangdiwala's $B$ underestimated agreement and overestimated agreement in the first two contingency tables. Normalized McNemar's~$\chi^2$ could not detect disagreement and agreement in second and third columns.
\end{itemize}

\begin{table}[H]
\caption{test performance of estimators with 2x2 extreme challenge tables showing situations of agreement or disagreement, but having unitary or null values in some cells (boldface showing discrepancies among tests).}
\label{tab:extremes}
\centering
\resizebox{\linewidth}{!}{
\fontsize{7}{9}\selectfont
\begin{tabular}[t]{lccccccccc}
\toprule
 & 	 & 	agr. c=1 & 	dis. d=1 & 	agr. b=1, c=1 & 	agr. b=0, c=1 & 	agr. d=0 & 	dis. d=0 & 	agr. c=0, d=0 & dis. c=0, d=0 \\ 
\midrule
 & & $\left[ \begin{array}{cc} 94 & 11  \\ 1 & 94 \end{array}\right]$ &  $\left[ \begin{array}{cc} 11 & 94  \\ 94 & 1 \end{array}\right]$ & $\left[ \begin{array}{cc} 99 & 1  \\ 1 & 99 \end{array}\right]$ & $\left[ \begin{array}{cc} 100 & 0  \\ 1 & 99 \end{array}\right]$ & $\left[ \begin{array}{cc} 180 & 10  \\ 10 & 0 \end{array}\right]$ & $\left[ \begin{array}{cc} 10 & 180  \\ 10 & 0 \end{array}\right]$ & $\left[ \begin{array}{cc} 190 & 10  \\ 0 & 0 \end{array}\right]$ & $\left[ \begin{array}{cc} 10 & 190  \\ 0 & 0 \end{array}\right]$  \\
\midrule
Holley and Guilford's $G$ & 	0/9 & 	0.88000 & 	-0.88000 & 	0.98000 & 	0.99000 & 	0.80000 & 	-0.90000 & 	0.90000 & 	-0.90000 \\ 
Gwet's $AC1$ & 	0/8 & 	0.88000 & 	-0.87531 & 	0.98000 & 	0.99000 & 	0.88950 & 	-0.89526 & 	0.94744 & 	-0.89526 \\ 
Scott's $\pi$ & 	2/8 & 	0.88000 & 	-0.88471 & 	0.98000 & 	0.99000 & 	\textbf{-0.05263} & 	-0.90476 & 	\textbf{-0.02564} & 	-0.90476 \\ 
Cohen's $\kappa$ & 	4/8 & 	0.88030 & 	-0.88471 & 	0.98000 & 	0.99000 & 	\textbf{-0.05263} & 	\textbf{-0.10465} & 	\textbf{0.00000} & 	\textbf{0.00000} \\ 
Corrected Cohen's $\kappa$  & 	4/8 & 	0.90025 & 	-0.88471 & 	1.00000 & 	0.99000 & 	\textbf{-0.05263} & 	\textbf{-0.10465} & 	\textbf{0.00000} & 	\textbf{0.00000} \\ 
Pearson's $r$ & 	4/8 & 	0.88471 & 	-0.88471 & 	0.98000 & 	0.99005 & 	\textbf{-0.05263} & 	\textbf{-0.68825} & 	\textbf{div/0} & 	\textbf{div/0} \\ 
Yule's $Q$ & 	6/8 & 	\textbf{0.99751} & 	\textbf{-0.99751} & 	0.99980 & 	1.00000 & 	\textbf{-1.00000} & 	\textbf{-1.00000} & 	\textbf{div/0} & 	\textbf{div/0} \\ 
Yule's $Y$ & 	6/8 & 	\textbf{0.93184} & 	\textbf{-0.93184} & 	0.98000 & 	1.00000 & 	\textbf{-1.00000} & 	\textbf{-1.00000} & 	\textbf{div/0} & 	\textbf{div/0}  \\ 
Shankar and Bangdiwala's $B$, & 	 & 	0.88581 & 	0.00608 & 	0.98010 & 	0.99005 & 	0.89503 & 	0.01786 & 	0.95000 & 0.05000 \\ 
verified by adjusted $B$ & 	3/8 & \textbf{0.77163} & \textbf{-0.98783} & 	0.96020 & 	0.98010 & 	0.79006 & 	\textbf{-0.96429} & 	0.90000 & 	-0.90000  \\
Dice's $F1$, & 	 & 	0.94000 & 	0.10476 & 	0.99000 & 	0.99502 & 	0.94737 & 	0.09524 & 	0.97436 & 	0.09524 \\ 
verified by adjusted $F1$ & 	0/8 & 	0.88000 & 	-0.79048 & 	0.98000 & 	0.99005 & 	0.89474 & 	-0.80952 & 	0.94872 & 	-0.80952 \\ 
Normalized McNemar's $\chi^2$ & 	5/8 & 	0.83333 & 	\textbf{0.00000} & 	\textbf{0.00000} & 	1.00000 & 	\textbf{0.00000} & 	0.89474 & 	\textbf{1.00000} & 	\textbf{1.00000}  \\ 
\bottomrule
\end{tabular}}
\end{table}

Based on this preliminary analysis, the famous Cohen's $\kappa$ failed in
most extreme situations. Other indices showed over and underestimation, were
unable to cope with disagreement, or failed to generate a coherent
value. The best estimators seem to be Holley and Guilford's~$G$ and Gwet's $AC1$. Scott's~$\pi$ and Dice's $F1$ are also competitive (since the rescalling of $F1$ makes possible the comparison with other coefficients).

\subsection{Inferential statistics~-~tables with $n=64$}\label{sec:n64}

We computed inferential statistics for all proposed estimators applying R functions from selected packages. When there was no function available, confidence interval was computed by bootstrapping~(described for the simple agreement coefficient,~$SAC$, ``\nameref{ap:G}'' in supplemental material). Exhaustive testing showed that Holley and Guilford's $G$, among all the studied estimators, minimized the discordance of inferential decisions from the others, and was selected as benchmark~(see~``\nameref{ap:mistakes}'' in supplemental material for details). 

Figure~\ref{fig:densityplots64}A shows that Holley and Guilford's $G$ is perfectly correlated with the proportion $\frac{a+d}{n}$, thus representing the bisectrix of reference, which is another evidence that $G$ can be a good choice for benchmark. The interval proportion ${0.391 \le \frac{a+d}{n} \le 0.609}$ corresponds to the non-rejection of the null hypothesis, $H_0:G=0$, interpreted here as populational neutrality (neither disagreement or agreement). The other two regions, denoted as H1- and H1+, correspond to the rejection of the null hipothesis, respectively meaning disagreement or agreement between raters. This interval appears in the subsequent panels~(Figures~\ref{fig:densityplots64}B~to~L), showing density~plots from the occurrences of tables (a total of 47,905~possible tables with $n=64$) for which the inferential decision coincided (marked as `correct', dashed lines) or was discrepant (denoted as `mistakes', solid lines) with the inferential test~of~$G$~(see supplemental material,~``\nameref{ap:G}''). In addition, some tables failed to compute due to invalid mathematical operations (e.g., division by zero) and, for some others, $p$ value could not be computed by particularities of their statistical calculation. 

Figure~\ref{fig:densityplots64} presents the estimators in order of total mistakes. It is interesting to realize that the summation of the density plots represented by the correct and mistaken decisions formed a virtually equal ogival shape, despite the different equations underlying each estimator. 

The small discordance between $SAC$ and $G$~(Figure~\ref{fig:densityplots64}B) is caused by differences between the bootstrapping and asymptotic statistical test (see supplemental material,~``\nameref{ap:G}''); for this reason a small amount of mistakes are located in the transition from H0 to H1 areas. Besides the amount of total mistakes~(solid lines in Figure~\ref{fig:densityplots64}), its location is also important:\mynobreakpar 

\begin{itemize}
	\item Gwet's~$AC1$ has no mistakes in H1+; mistakes in H1- are close to the transition to H0~(Figure~\ref{fig:densityplots64}C).
	\item Scott's~$\pi$ does the reverse, with no mistakes in H1-~(Figure~\ref{fig:densityplots64}D).
	\item Cohen's~$\kappa$~(Figure~\ref{fig:densityplots64}E), Pearson's~$r$~(Figure~\ref{fig:densityplots64}F), and Yule's~$Q$~(Figure~\ref{fig:densityplots64}H) are similar; the first two presented similar percentages of mistakes in all three areas, while the latter had more mistakes when rejecting the null hypothesis.
	\item Yule's~$Y$~(Figure~\ref{fig:densityplots64}G) was slightly better than Yule's~$Q$. 
	\item The original Shankar and Bangdiwala's~$B$~(Figure~\ref{fig:densityplots64}J) was mistaken in all situations of neutrality (rejecting the null hypothesis); also, when disagreement between raters was high, it also had mistaken decisions (assuming neutrality). Paradoxically, it showed perfect performance for H1+ region.
	\item Our proposition of adjustment by rescalling Shankar and Bangdiwala's~$B$~(Figure~\ref{fig:densityplots64}I) slightly improved the total number of mistakes, but created a mixing situation under H0 and displaced the mistakes to H1+.
	\item Dice's~$F1$~(Figure~\ref{fig:densityplots64}K) and Normalized McNemar's~$\chi^2$~(Figure~\ref{fig:densityplots64}L) produced a flawed approach to the inferential statistics: Dice's~$F1$ has the majority of mistakes in the H0~area; Normalized McNemar's~$\chi^2$ not only had more mistakes in H0, but also more mistakes than correct decisions when the agreement between raters is high (right of H1+~area). 
\end{itemize}

\begin{figure}[H]
\begin{center}
\includegraphics[width=0.75\textwidth]{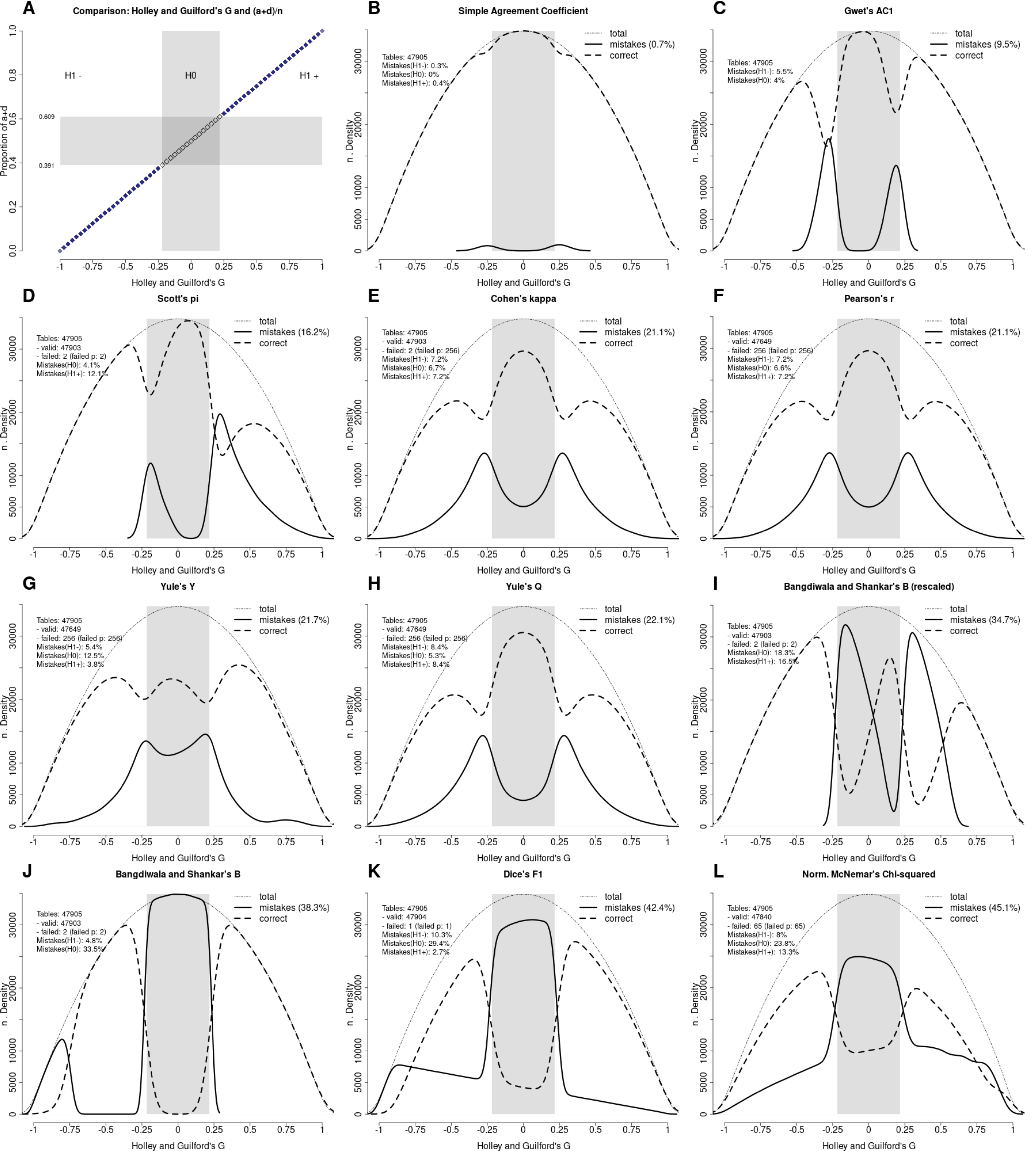}
\end{center}
\caption{Relative perfomance of proposed estimators in relation to inferential statistics of Holley and Guilford's coefficient ($G$). $G$ is a linear transformation of $(a+d)/n$. Gray area corresponds to $p \ge 0.05$. Mistakes are discordance of inferential decision and fails are impossibility of estimator computation (e.g., division by zero) provided as the proportion in relation to all 47,905 possible tables with $n=64$. }\label{fig:densityplots64}
\end{figure}

\subsection{Comprehensive maps~-~all tables with $1 \le n \le 68$}\label{sec:n1to68}

Tables with size ranging from 1 to 68 were generated (see supplemental material,~``\nameref{ap:n1to68}''), which resulted in a little more than one million different 2x2~tables covering all possible arrangements of `$abcd$'. 

A global measurement of estimator qualities was computed by the Pearson's and Spearman's correlations between $G$ and all other estimators across all 1,028,789 tables~(Table~\ref{tab:correlation}). Pearson's correlation (not to be confounded with Pearson's $r$ application to aggrement under investigation here) assesses linear trend, while Spearman's assesses monotonic trend of each pair of estimators.

\begin{table}[H]
\centering
\caption{Pearson's and Spearman's correlations coefficients between Holley and Guilford's~$G$ and all the other estimators: correlations were separately estimated for each $n$ (from 1 to 68) and, then, lower~(HDI LB) and upper~(HDI UB) bounds of the 95\% highest density interval were obtained (see supplemental material,~``\nameref{ap:r_hdi}''). Table rows ordered by median of Spearman's correlations.}\label{tab:correlation}
\hspace{0.3in}
\resizebox{\linewidth}{!}{
\begin{tabular}{l|ccc|ccc}
\toprule
                                 & \multicolumn{3}{c}{\textbf{Pearson}}                & \multicolumn{3}{c}{\textbf{Spearman}}               \\
\textbf{Estimator}               & \textbf{Median} & \textbf{HDI LB} & \textbf{HDI UB} & \textbf{Median} & \textbf{HDI LB} & \textbf{HDI UB} \\
\midrule
Gwet's $AC1$                             & 0.9931 & 0.9923 & 0.9934 & 0.9933 &  0.9899  & 0.9943 \\
Shankar and Bangdiwala's $B$             & 0.9698 & 0.9677 & 0.9713 & 0.9772 &  0.6699  & 0.9890 \\
Adjusted $B$                             & 0.9698 & 0.9677 & 0.9713 & 0.9772 &  0.6699  & 0.9890 \\
Scott's $\pi$                            & 0.9555 & 0.9315 & 0.9643 & 0.9578 &  0.9385  & 0.9662 \\
Pearson's $r$                            & 0.9131 & 0.9089 & 0.9474 & 0.8661 &  0.3033  & 0.9583 \\
Cohen's kappa                            & 0.8713 & 0.7973 & 0.8928 & 0.8659 &  0.7925  & 0.8897 \\
Cohen's $\kappa$ corrected by $\kappa_M$ & 0.8351 & 0.7770 & 0.8596 & 0.8371 &  0.7775  & 0.8604 \\
Dice's $F1$                              & 0.7665 & 0.7349 & 0.7792 & 0.7611 &  0.7378  & 0.7751 \\
Adjusted $F1$                            & 0.7665 & 0.7349 & 0.7792 & 0.7611 &  0.7378  & 0.7751 \\
Yule's $Q$                               & 0.7841 & 0.7147 & 0.8326 & 0.7182 &  0.2305  & 0.8818 \\
Yule's $Y$                               & 0.7384 & 0.6704 & 0.8000 & 0.7182 &  0.2305  & 0.8818 \\
Normalized McNemar's $\chi^2$            & 0.0968 & 0.0084 & 0.3324 & 0.1089 & -0.0316  & 0.6615 \\
Traditional McNemar's $\chi^2$           & -0.3978 & -0.4202 & -0.3126 & -0.3066 & -0.3950  & 0.2880 \\
\bottomrule
\end{tabular}}
\end{table}

As defined in the previous section, Holley and Guilford's~$G$ was adopted as benchmark. It is possible to observe that Gwet's~$AC1$ has the best correlation with Holley and Guilford's~$G$, but many others also show acceptable correlations. However, the correlation was lower for Cohen's~$\kappa$, Yule's~$Q$, Dice's~$F1$, and Yule's~$Y$, and much lower, close to or absent, for Normalized and Traditional McNemar's~$\chi^2$.

\begin{figure}[H]
\begin{center}
\includegraphics[width=0.75\textwidth]{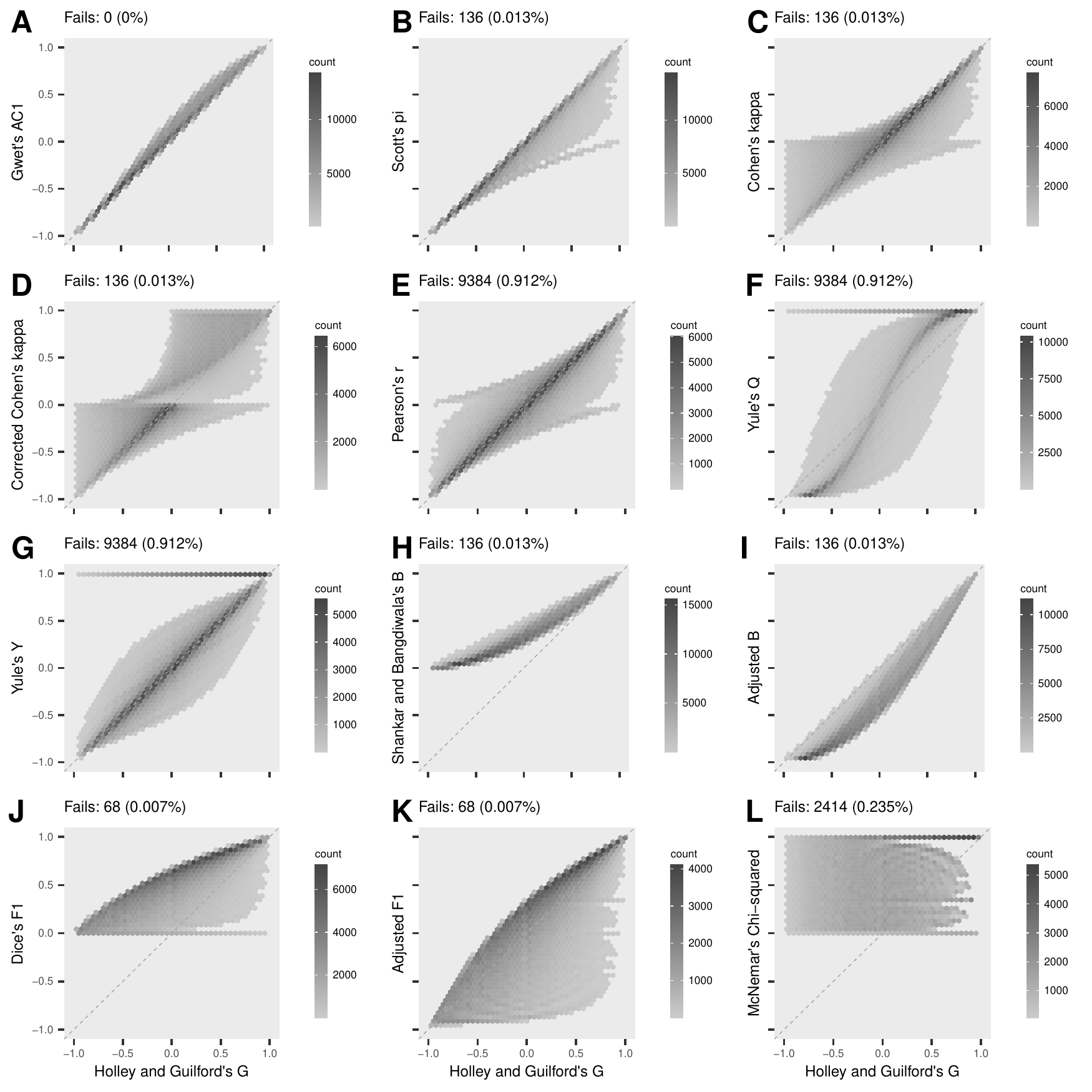}
\end{center}
\caption{Hexbin plots of estimator computation of all possible 2x2~tables with $n$ ranging from 1 ot 68 (total of 1,028,789~different tables; percentage is of non computable tables). Holley and Guilford's~$G$ was adopted as benchmark. Darker hexbins correspond to the location of higher counting of tables: good estimators should have darker hexbins and lower dispersion around the bisector line.}\label{fig:hexbin1to68}
\end{figure}

A more detailed view to assess the quality of each estimator is in Figure~\ref{fig:hexbin1to68}. According to this second criteria, again the best estimator is Gwet's~$AC1$~(Figure~\ref{fig:hexbin1to68}A), and the worst is Normalized McNemar's~$\chi^2$~(Figure~\ref{fig:hexbin1to68}L). The Traditional McNemar's~$\chi^2$ is not comparable to the other estimators because it does not provide values in the interval [-1,1]~(although not shown here, its mapping was tested with procedures available in the supplemental material). Cohen's~$\kappa$~(Figure~\ref{fig:hexbin1to68}C), Pearson's~$r$~(Figure~\ref{fig:hexbin1to68}E), Yule's~$Q$~(Figure~\ref{fig:hexbin1to68}F)~and~$Y$~(Figure~\ref{fig:hexbin1to68}G), and Dice's~$F1$~(Figure~\ref{fig:hexbin1to68}J) are mediocre estimators of agreement. Rescalled~$F1$~(Figure~\ref{fig:hexbin1to68}K) aligned its darker hexbins with the bisector, but could not fix the number of tables with mistaken estimatives below the bisector.

Shankar and Bangdiwala's~$B$~(in its original form,~Figure~\ref{fig:hexbin1to68}H) is defective, with darker hexbins close to the bisector line only when $G$ approaches 1. When scaled to [-1,1]~(Figure~\ref{fig:hexbin1to68}I) the alignment is improved but it leaves darker hexbins away and ligher hexbins close to the bisector line. From table~\ref{tab:correlation}, one should expect a better performance of $B$. In this figure it is possible to observe that its excellent correlation depended on fairly aligned pairs of $G$ and $B$ observations but, in terms of linear regression, the great number of tables that are not close to the bisectrix leads to the not so good performance observed in~Figure~\ref{fig:densityplots64}. It is to say that such a simple rescalling of $B$ cannot fix this estimator and it is structurally defective.

\begin{figure}[H]
\begin{center}
\includegraphics[width=0.75\textwidth]{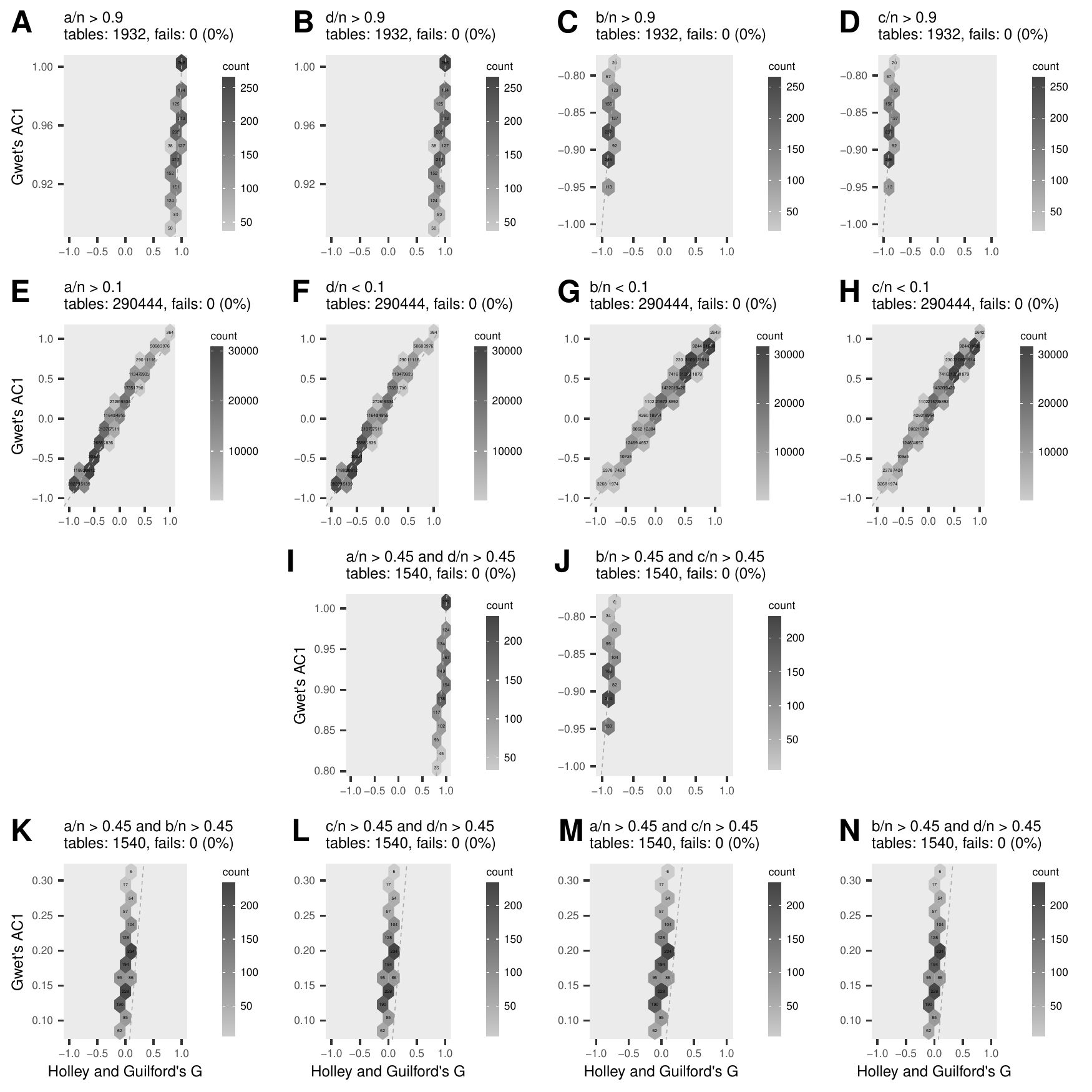}
\end{center}
\caption{Hexbin plots of Gwet's~$AC1$ computation of extreme 2x2~tables with $n$ ranging from 1 ot 68 (total of 1,028,789~different tables; percentage is of non computable tables) in comparison to Holley and Guilford's~$G$ (adopted as benchmark). First row (A to D): each cell contains more than 90\% of all data. Second row (E to H): each cell contains less than 10\% of all data. I: 90\% of data in main diagonal; J: 90\% of data in off-diagonal. Forth row (K to N): respectively with 90\% of data in the first row, second row, first columns, and second column of a 2x2~table. Scale of $y$-axis varies to show the behavior of coincidences between the estimators. Dashed gray line represents the bisectrix.}\label{fig:hexbin_AC1}
\end{figure}

\begin{figure}[H]
\begin{center}
\includegraphics[width=0.75\textwidth]{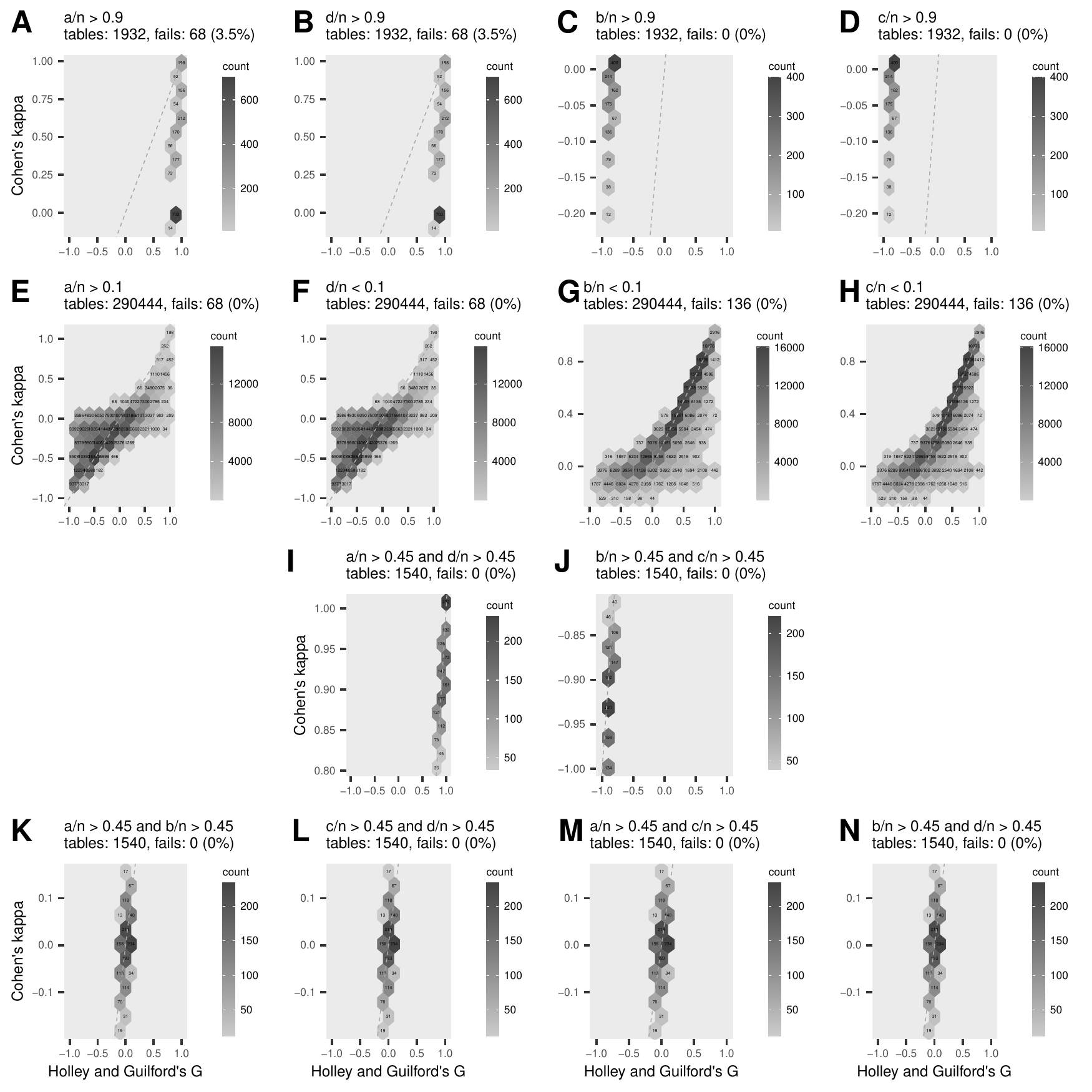}
\end{center}
\caption{Hexbin plots of Cohen's~$\kappa$ computation of extreme 2x2~tables with $n$ ranging from 1 ot 68 (total of 1,028,789~different tables; percentage is of non computable tables) in comparison to Holley and Guilford's~$G$ (adopted as benchmark). First row (A to D): each cell contains more than 90\% of all data. Second row (E to H): each cell contains less than 10\% of all data. I: 90\% of data in main diagonal; J: 90\% of data in off-diagonal. Forth row (K to N): respectively with 90\% of data in the first row, second row, first columns, and second column of a 2x2~table. Scale of $y$-axis varies to show the behavior of coincidences between the estimators. Dashed gray line represents the bisectrix.}\label{fig:hexbin_k}
\end{figure}

\begin{figure}[H]
\begin{center}
\includegraphics[width=0.75\textwidth]{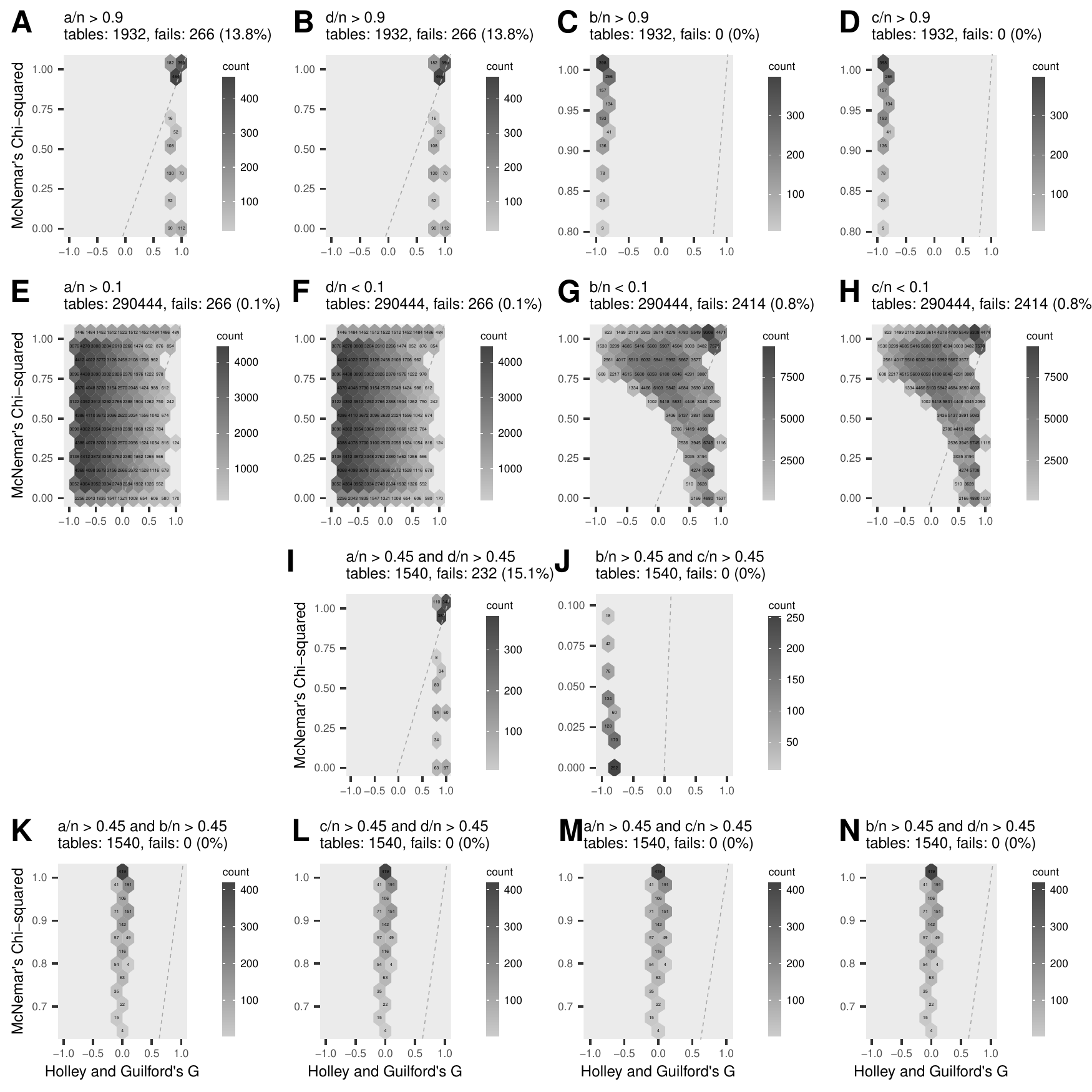}
\end{center}
\caption{Hexbin plots of Normalized McNemar's~$\chi^2$ computation of extreme 2x2~tables with $n$ ranging from 1 ot 68 (total of 1,028,789~different tables; percentage is of non computable tables) in comparison to Holley and Guilford's~$G$ (adopted as benchmark). First row (A to D): each cell contains more than 90\% of all data. Second row (E to H): each cell contains less than 10\% of all data. I: 90\% of data in main diagonal; J: 90\% of data in off-diagonal. Forth row (K to N): respectively with 90\% of data in the first row, second row, first columns, and second column of a 2x2~table. Scale of $y$-axis varies to show the behavior of coincidences between the estimators. Dashed gray line represents the bisectrix.}\label{fig:hexbin_MN}
\end{figure}

In case of more extreme 2x2~tables, the behavior of Gwet's~$AC1$~(the estimator that better captured the estimatives by $G$), Cohen's~$\kappa$~(the most popular coefficient of agreement), and Normalized McNemar's~$\chi^2$~(also popular but, again, the less reliable agreement estimator according to our analysis) are presented in figures~\ref{fig:hexbin_AC1}, \ref{fig:hexbin_k}, and \ref{fig:hexbin_MN}. 

\subsection{Replication of this research}\label{sec:replication}

All procedures presented in this work, and some others not shown in the main text, can be replicated with R scripts, available in\newline\newline
\verb"https://sourceforge.net/projects/tables2x2"\newline\newline
Instructions are in \newline
\verb"Kappa_README.pdf"\newline
containing:\mynobreakpar
\begin{itemize}
\item \ref{ap:Rfunctions}~(``\nameref{ap:Rfunctions}") describes the implementation of all mentioned estimators, including their inferential statistic tests. 
\item \ref{ap:challenge}~(``\nameref{ap:challenge}") replicates findings from Tables~\ref{tab:challenge} and \ref{tab:extremes}.
\item \ref{ap:r_hdi}~(``\nameref{ap:r_hdi}") shows the procedure to replicate Table~\ref{tab:correlation} and some additional figures.
\item \ref{ap:n1to68}~(``\nameref{ap:n1to68}") has the procedures to generate all possible tables, given any range of $n$.
\item \ref{ap:n64}~(``\nameref{ap:n64}") exemplifies with creation of all 47,905 possible tables with $n=64$ to build Figure~\ref{fig:densityplots64}.
\item \ref{ap:fig_hexbin}~(``\nameref{ap:fig_hexbin}") shows how to create 1,028,789 possible tables with $1 \le n \le 68$ and provide the R code to generate hexbin Figures~\ref{fig:hexbin1to68},~\ref{fig:hexbin_AC1},~\ref{fig:hexbin_k},~and~\ref{fig:hexbin_MN}.
\end{itemize}

\section{Discussion}\label{sec:discussion}

Agreement coefficients are fundamental statistics. Except for  Francis Galton and Karl Pearson starting in the 1880s, who created the correlation coefficient (without the intention to apply it to agreement, although the computation of this work shows its equivalence) and the pioneer work of Yule, 1912~\cite{Yule1912} (who created $Q$ and $Y$ coefficients, expliciting intending to measure association between nominal variables), there was great interest and creation of agreement coefficients between the 1940s and 1970s~(Dice, 1945~\cite{Dice1945}; Cramér, 1946~\cite{Cramer1946}; McNemar, 1947~\cite{McNemar1947}; Scott, 1955~\cite{Scott1955}; Cohen, 1960~\cite{Cohen1960}; Holley \& Guilford, 1964~\cite{Holley1964}; Matthews, 1975~\cite{Matthews1975}; Hubert, 1977~\cite{Hubert1977}), including most of the coefficients in use today. Renewed interest appears in this century, with the creation of new coefficients, searching for improvement and avoidance of known flaws of the older propositions~(Gwet, 2008 and 2010~\cite{Gwet2008,Gwet2010}; Shankar \& Bangdiwala, 2014~\cite{Shankar2014}), or attempts to improve traditional estimators~(e.g., Lu, 2010~\cite{Lu2010}, Lu et al., 2017~\cite{Lu2017}).

Although not in the chronological order, the two most used coefficients explored here are McNemar's~$\chi^2$, praised in Epidemiology textbooks~\cite{Kirkwood2003}(p.~218), and the widespread and famous Cohen's~$\kappa$, which was the departure point of the present work. The first contingency table presented in the seminal paper of Cohen, 1960~\cite{Cohen1960} shows two clinical psychologists classifying individuals in schizophrenic, neurotic or brain-damaged categories. This table is an example of a 3x3~table with proportions of agreement or disagreement between raters, while researchers more often pursue agreement in 2x2~tables. It follows a lengthy discussion of traditional measures of agreement such as Pearson's~$\chi^2$ and contingent coefficient. Curiously, for this first contingency table, Cohen only computed $\chi^2$ and reported it as significant, concluding by the existence of association but arguing that $\chi^2$ is not a defensible measurement of agreement. However, in 2x2~tables both association and agreement are coincident and provide equal $p$~values, Pearson's~$\chi^2$ and Cohen's~$\kappa$ are equivalent~\cite{Feingold1992}, which may suggest that, at least in 2x2~tables, Cohen's~$\kappa$ is a mere test of association.  The computation of $\kappa$ is not presented for this first table, thus we computed~$\kappa=-0.092$, concluding that this is an example of slight disagreement between the raters, an unfortunate initial example for who is presenting a new measure of agreement. It follows the introduction of $\kappa$ calculation, mentioning other similar statistics such as Scott's~$pi$, which was published in 1955~\cite{Scott1955}. When a second contingency table is presented, with comparison of the calculation of $\chi^2$ and $\kappa$, it brings the same storyline of psychologists classifying patients in three categories, but the numbers are changed, now creating a scenario of rater's agreement! In addition, Cohen proposes the correction of $\kappa$ by the maximum $kappa$, which was largely forgoten in the literature~(presented as $k_M$ in ``Methods,~\nameref{sec:cohens-kappa}") only for positive estimation of $\kappa$. For the lower limit of $\kappa$ we quote:\mynobreakpar
\begin{quote}
The lower limit of K is more complicated, since it depends on the marginal distributions. [...] Since $\kappa$ is used as a measure of agreement, the complexities of its lower limit are of primarily academic interest. It is of importance that its upper limit be 1.00. If it is less than zero (i.e., if the observed agreement is less than expected by chance), it is likely to be of no further practical interest.\newline Cohen, 1960~\cite{Cohen1960}
\end{quote}

From this consideration we have decided not to attempt any correction for negative values of $\kappa$.

Part of the problem is to define what are really being measured by any of these coefficients. Lienert, 1972~\cite{Lienert1972} argued that tests for independence are not from the same nature of tests for agreement. Independence tests the null hypothesis 
$$H_{0,~ind}:(\alpha \cdot \delta) - (\beta \cdot \gamma) = 0$$
while agreement assesses 
$$H_{0,~agr}:(\alpha+\delta)-(\beta+\gamma) = 0$$
 where $\alpha$, $\beta$, $\gamma$, and $\delta$ are the populational proportions respectively estimated by $a$, $b$, $c$, and $d$ (see notation on Table~\ref{tab:mathnotation}). Independence tests (from which Cohen's $\kappa$, Pearson's $\phi$ and other Pearson's $\chi^2$-based statistics are representatives), and agreement tests (from which $G$ is a representative) are, therefore, sensitive to different types of association~\cite{Shreiner1980}.
 
In the same line of reasoning, Pearson's~$\chi^2$ and contingent coefficient, as well McNemar test~\cite{McNemar1947}, were previously criticized by Cohen, 1968~\cite{Cohen1968} who stated that association does not imply necessarily in agreement for any table size. This aspect will be further discussed below. The question is to know how these coefficients are related, when they measure association or agreement, and when their measures are coincident or discrepant.

Holley \& Guilford, 1964~\cite{Holley1964} showed that $G$ is equal to Pearson's $\phi$ coefficient only when the marginal values, i.e., $\frac{a+b}{n} = \frac{a+c}{n} = \frac{b+d}{n} = \frac{c+d}{n} = 0.5$ where $n = a+b+c+d$, a condition in which $G=\phi=0$ but also $\kappa=0$.
It is to say that $\kappa$ and $\phi$ are, otherwise, different entities of $G$, with potential different performances to detect independence or agreement in 2x2~contingency tables: $\phi$ and $G$ are related~\cite{Holley1964} by\mynobreakpar

\begin{eqnarray}
\label{eq:Gphi}
\begin{aligned}[b]
\phi = \frac{ \frac{G}{4} - \left( \frac{a+b}{n} - \frac{1}{2} \right) \left( \frac{a+c}{n} - \frac{1}{2} \right)  }{ \sqrt{\frac{a+b}{n} ~ \frac{a+c}{n} ~ \frac{b+d}{n} ~ \frac{c+d}{n}} }
\end{aligned}
\end{eqnarray}

and $\kappa$ is related to $G$~\cite{Green1981} by\mynobreakpar
 
\begin{eqnarray}
\label{eq:Gkappa}
\begin{aligned}[b]
\kappa = {\frac{ \frac{G+1}{2} - p_c }{1-p_c}}
\end{aligned}
\end{eqnarray}

The parcel $p_c$, included in the computation of $\kappa$, is known in the literature as `chance correction factor'. Since $\kappa$ includes more parcels, one would expect that the performance of $\kappa$ should exceed that of $G$. However, it was stated by Green, 1981~\cite{Green1981} that ``Because the standard equation for kappa clearly includes a chance correction factor, many authors [...] have suggested its usage. Unfortunately, chance has not been explicitly defined." It was also shown that, under skewed marginals, $\kappa$ and $\phi$ underestimate agreement, while $G$ is a stable estimator~\cite{Shreiner1980}.

It seems that it was long understood by theoreticians that $G$ is a superior estimator to Cohen's~$\kappa$, but practitioners of applied statistics, for some reason, adhered to the latter. These theorectical reasons, in addition to our findings, leaded to the use of the Holley and Guilford's $G$ index as a reference to the performance of concurrent estimators. 
 
In fact, many authors compare several of the estimators but, to our knowledge, no one performed an exhaustive analysis with hundreds of thousands tables as presented here to obtain a comprehensive map of estimator behaviors. Many comparisons stick to some particular cases, using challenging tables similar to that in Tables~\ref{tab:challenge} and \ref{tab:extremes}, sometimes to show weakeness or strengh of particular coefficients in particular situations. Even so, many of their conclusions point to many Cohen's $\kappa$ problems and favor Holley and Guilford's $G$~\cite{Shreiner1980,Green1981} or Gwet's~$AC1$~\cite{Kuppens2011,Wongpakaran2013,Xie2017}.

Accordingly to our results, by assuming Holley and Guilford's $G$ as benchmark, Gwet's~$AC1$ is also a good estimator. Not only $AC1$ mistakes are the lowest in inferential statistics, but also these mistakes are located around the null hypothesis, which is ---~paraphasing~Cohen, 1960~\cite{Cohen1960}~--- ``unlikely in practice"~(Figure~\ref{fig:densityplots64}C). $AC1$ agreement is also close to the bisector line (darker hexbins in Figure~\ref{fig:hexbin1to68}A) and it is not confounded by extreme tables (Figure~\ref{fig:hexbin_AC1}).

Cohen's~$\kappa$ is not only a poor estimator of agreement, but also prone to provide incorrect statistical decisions when there is neutrality and not large disagreement or agreement between raters. Figure~\ref{fig:densityplots64}E shows that it is mistaken in around 21\% of all possible tables with $n=64$, with roughly one third in each region, but this distribution is not uniform: mistakes are less likely to occur away of the greater disagreements or agreements. Another weakness is that there were problems in 256 tables of size $n=64$. Two tables failed when all data are in $a$ or $d$ (a clear 100\% agreement) because it causes a division by zero (equation~\ref{eq:kappa_abcd}). In addition to that, the inferential decision provided by \verb"epiR::epi.kappa" ($z$ statistic for $\kappa$ associated with $p$~value) is also unavailable in other 254 tables when one row or column is empty (i.e., $a+b=0$ or $a+c=0$ or $b+d=0$ or $c+d=0$). Finaly, Figure~\ref{fig:hexbin_k} shows some details of $\kappa$ limitations. In scenarios of agreement in which most of data are concentraded in $a$ or $d$ with $\kappa$ mistakenly providing values from 0 to 1 (mostly 0, Figures~\ref{fig:hexbin_k}A~and~B). That happens due to equation~\ref{eq:kappa_abcd}, for $a$ and $d$ appear in both parcels of the denominator creating an exagerated value while the parcel $ad$ in the numerator is a low value, thus $\kappa$ underestimates the agreement in these situations. Concentration in $b$ or $c$, which appear only one in each denominator parcel, only causes a small number in the numerator due to $bc$~(always a small number due to high~$b$ and low~$c$ or vice-versa), leading to underestimation of the disagreement~(Figures~\ref{fig:hexbin_k}C~and~D). Cohen's~$\kappa$ also has problems one only one of the four cells is relatively empty, which may happen, for instance, if one of the raters is more lenient than the other; in these cases Cohen's~$\kappa$ assessment has excess of variability~(Figures~\ref{fig:hexbin_k}E~to~H). A last comment on Cohen's~$\kappa$ is that the correction by maximum $~\kappa$, proposed by Cohen himself and forgoten by following researchers, may be a correction for effect size~(Table~\ref{tab:kappacategory}) but it seems to worse the general estimator behavior.

We started this review by criticizing Cohen's~$\kappa$ but the investigation of alternatives leaded, in the end, to the development of a method to assess any other estimator, some of them described along this text. For that, McNemar's~$\chi^2$ was a collateral damage. It was not anticipated such a poor performance for such a widespreaded estimator. 
On the other hand, we tested the performance of McNemar's~$\chi^2$ to verify if it could be applied to a general situation of agreement, concluding that it cannot be used. To be fair, one must recognize that McNemar's~$\chi^2$ test was designed to a very specific situation: change of signal in a pre-post scenario, using only the off-diagonal~\cite{McNemar1947}. Unfortunately, McNemar's~$\chi^2$ sometimes is applied in the context of agreement between methods~(e.g.,~Kirkwood and Sterne, 2003,~pp.~216-218~\cite{Kirkwood2003}). By using only $b$ and $c$, it becomes a futile mental exercise to link rejection of non-rejection of the null hypothesis with agreement or disagreement between raters. For instance, in trying to interpret McNemar's measure as agreement between raters, $b=26, c=10$ leads to $MN=0.44, IC95\%(MN)=[0.14,0.74]$, rejects $H_0$ and would suggest disagreement, but $b=18, c=18$ (which is the same amount of disagreement) leads to $MN=0.00, IC95\%(MN)=[-0.02, 0.34]$, does not reject $H_0$ and would suggest agreement. Both decisions completly disregard the agreement values, $a$ and $d$, it does not matter if they are 2 or 2 thousand. It would only matters that one rater sistematically opposes the other and that one of them is biased to provide much more positives or negatives than the other. It they are perfectly opposed in provide assessment to make $b=c$, then this perfect disagreement would be no more detectable. Consequently, the application of McNemar's~$\chi^2$ as a measure of association or agreement, removing it from its original context, can only lead to the confusing results observed on the challenge 2x2~tables with inability to detect agreement or disagreement in some situations and overestimation of agreement or disagreement in others~(Tables~\ref{tab:challenge} and \ref{tab:extremes}), fails in coincidence of the null hypothesis~(Figure~\ref{fig:densityplots64}L), and mixed estimatives along clear situations of agreement or disagreement~(Figure~\ref{fig:hexbin1to68}L). In essence, McNemar's~$\chi^2$ is not an agreement estimator and its use must be restricted to its original context.

To close this discussion, a brief comment on the other estimators is in order. Scott's~$\pi$ appears in third place. It is not a bad estimator but, contrary to its original proposition as inter-rater reliability, it seems a more reliable estimator of disagreement~(Figure~\ref{fig:densityplots64}D). Another way to confirm this statement is to observe that most of its correct countings are on the bisectrix, but the dispersion increases with increasing rater agreement~(Figure~\ref{fig:hexbin1to68}K).

Pearson's~$r$ is, primarily, a measure of association. However, in 2x2~tables its performance is also very similar to Cohen's~$\kappa$, conceived to measure agreement, which can be observed by similar density plots (Figures~\ref{fig:densityplots64}E~and~F), mapping of hexbins~(Figures~\ref{fig:hexbin1to68}C~and~E) or their similar correlations with $G$ in Table~\ref{tab:correlation}. We observe that equations for $\kappa$~(equation~\ref{eq:kappa_abcd}), $Q$~(equation~\ref{eq:Q}), $r$~(equation~\ref{eq:r}, which is also equal to $\rho$, $\tau$, $\phi$, and Cramér's $V$), all have a sort of $OR$~(equation~\ref{eq:OR}) in their numerators~($ad-bc$). Although having similar global performances, the deficiencies of $r$ and $\kappa$ are due to different reasons (see Tables~\ref{tab:challenge}~and~\ref{tab:extremes}). 

Yule's~$Q$ and $Y$ come next, with small advantage to $Y$. Their global performance are close to that of Cohen's~$\kappa$ and Pearson's~$r$, but $Y$ made more mistakes when there is neutrality~(i.e., in the region of the non-rejection of the null hypothesis), while $Q$ made more mistakes when there are disagreement or agreement between raters~(i.e., in the regions of rejection of the null hypothesis, Figures~\ref{fig:densityplots64}G~and~H). Both also show a tendency to overestimate agreement providing values equal to 1 for any disagreement, neutrality or agreement provided by $G$, which is represented by the horizontal lines on the top of Figures~\ref{fig:hexbin1to68}F and \ref{fig:hexbin1to68}G, specially for higher agreements (darker hexbins); this also explains the results observed in Table~\ref{tab:extremes}. At least, as it happens to be with Scott's~$\pi$, Yule's~$Y$ concentrates most of their estimatives around the bisector line, while Yule's~$Q$ is more problematic, showing a sigmoid shadow of darker hexbins that explains its greater tendency do overestimate both agreement and disagreement as shown in Table~\ref{tab:challenge}.

Dice's~$F1$ and Shankar and Bangdiwala's~$B$, alike Normalized McNemar's~$\chi^2$, only provide positive estimatives, thus it is required to look at the 2x2~contingency table to decide if it is agreement or disagreement when the null hypothesis is rejected. Dice's~$F1$~(Figure~\ref{fig:hexbin1to68}J) has dispersed results. Its normalized version~(Figure~\ref{fig:hexbin1to68}K) as well as Normalized McNemar's~$\chi^2$ (Figure~\ref{fig:hexbin1to68}L) fills around half of the graph areas. Incidentaly, $F1$ also does not use the entire information from a 2x2~table, leaving $d$ out of reach. Perhaps, the first criteria to be a good agreement estimator should be to apply the whole information available. Shankar and Bangdiwala's~$B$ has more serious problems. It was the only estimator that could be mistaken (i.e., mistakenly rejecting) in all situations of null hypothesis~(Figure~\ref{fig:densityplots64}J), in addition to cases of higher disagreement (not rejecting neutrality). Like Gwet's $AC1$~(Figure~\ref{fig:densityplots64}C), it has no mistakes on the H1+ area~---~although the performance of $AC1$ is a lot better.

Since it is confusing to compare a range [0,1] with [-1,1], we propose the rescaling by $2i-1$ (where $i$ is $F1$ or $B$). It provided a partial fix for Dice's~$F1$ for it now placed most of the higher counts along the bisector, but it still shows a large shadow of uncertainty~(Figure~\ref{fig:hexbin1to68}K). Rescalling Shankar and Bangdiwala's~$B$ inverted its problem~(Figure~\ref{fig:densityplots64}I): the total amount of mistakes practically remains, but there is no mistakes in H1- area; now it is like Scott's~$\pi$~(Figure~\ref{fig:densityplots64}D) but, even so, Scott's~$\pi$ is a better choice. Improvement happened on the dispersion of estimatives along the bisector line, but it did not reflect a great improvement because the region with higher concentration of tables is away of the bisector (darker hexbins in~Figure~\ref{fig:hexbin1to68}I). 

This work has a humble mission for restoring Holley and Guilford's~$G$ as the best agreement estimator, closely followed by Gwet's~$AC1$. Both have inferential statistics associated to them, in order to satisfy research requirements. Gwet's~$AC1$ was already implemented in R packages. We could not find any Holley and Guilford's~$G$ implementation but the R scripts presented in supplemental material in the section named ``~\nameref{ap:G}'' can be easily adapted, including the asymptotic test proposed in the literature for tables with $n>30$ (bootstrapping techniques are easy to adapt for smaller tables). Holley and Guilford's~$G$ and Gwet's~$AC1$ should be considered by modern researchers as the first choices for agreement measurement in 2x2~tables.

\newpage
\bibliography{kappa} 

\begin{thebibliography}{}

\bibitem[Banerjee et~al., 1999]{Banerjee1999}
Banerjee, M., Capozzoli, M., McSweeney, L., and Sinha, D. (1999).
\newblock Beyond kappa: A review of interrater agreement measures.
\newblock {\em Canadian Journal of Statistics}, 27.

\bibitem[Cohen, 1960]{Cohen1960}
Cohen, J. (1960).
\newblock A coefficient of agreement for nominal scales. educational and
  psychological measurement.
\newblock {\em Educational and Psychological Measurement}, 20.

\bibitem[Cohen, 1968]{Cohen1968}
Cohen, J. (1968).
\newblock Weighted kappa: Nominal scale agreement provision for scaled
  disagreement or partial credit.
\newblock {\em Psychological Bulletin}, 70.

\bibitem[Cramér, 1946]{Cramer1946}
Cramér, H. (1946).
\newblock {\em Mathematical methods of statistics}.
\newblock Princeton University Press.

\bibitem[Dice, 1945]{Dice1945}
Dice, L.~R. (1945).
\newblock Measures of the amount of ecologic association between species.
\newblock {\em Ecology}, 26.

\bibitem[Efron, 2007]{Efron2007}
Efron, B. (2007).
\newblock Bootstrap methods: Another look at the jackknife.
\newblock {\em The Annals of Statistics}, 7.

\bibitem[Feingold, 1992]{Feingold1992}
Feingold, M. (1992).
\newblock The equivalence of cohen's kappa and pearson's chi-square statistics
  in the 2 × 2 table.
\newblock {\em Educational and Psychological Measurement}, 52.

\bibitem[Green, 1981]{Green1981}
Green, S.~B. (1981).
\newblock A comparison of three indexes of agreement between observers:
  Proportion of agreement, g-index, and kappa.
\newblock {\em Educational and Psychological Measurement}, 41.

\bibitem[Gwet, 2008]{Gwet2008}
Gwet, K.~L. (2008).
\newblock Computing inter-rater reliability and its variance in the presence of
  high agreement.
\newblock {\em British Journal of Mathematical and Statistical Psychology}, 61.

\bibitem[Gwet, 2010]{Gwet2010}
Gwet, K.~L. (2010).
\newblock {\em Handbook of Inter-Rater Reliability: the definitive guide to
  measuring the extent of agreement among raters}.
\newblock 3rd edition.

\bibitem[Hoff et~al., 1982]{Hoff1982}
Hoff, R., Hoff, R., Sleigh, A., Mott, K., Barreto, M., de~Paiva, T.~M.,
  de~Souza~Pedrosa, J., and Sherlock, I. (1982).
\newblock Comparison of filtration staining (bell) and thick smear (kato) for
  the detection and quantitation of schistosoma mansoni eggs in faeces.
\newblock {\em Transactions of the Royal Society of Tropical Medicine and
  Hygiene}, 76.

\bibitem[Holley and Guilford, 1964]{Holley1964}
Holley, J.~W. and Guilford, J.~P. (1964).
\newblock A note on the g index of agreement.
\newblock {\em Educational and Psychological Measurement}, 24.

\bibitem[Hripcsak and Rothschild, 2005]{Hripcsak2005}
Hripcsak, G. and Rothschild, A.~S. (2005).
\newblock Agreement, the f-measure, and reliability in information retrieval.
\newblock {\em Journal of the American Medical Informatics Association}, 12.

\bibitem[Hubert, 1977]{Hubert1977}
Hubert, L. (1977).
\newblock Nominal scale response agreement as a generalized correlation.
\newblock {\em British Journal of Mathematical and Statistical Psychology}, 30.

\bibitem[Janson and Vegelius, 1982]{Janson1982}
Janson, S. and Vegelius, J. (1982).
\newblock The j-index as a measure of nominal scale response agreement.
\newblock {\em Applied Psychological Measurement}, 6.

\bibitem[King et~al., 2012]{King2012}
King, N.~B., Harper, S., and Young, M.~E. (2012).
\newblock Use of relative and absolute effect measures in reporting health
  inequalities: Structured review.
\newblock {\em BMJ (Online)}, 345.

\bibitem[Kirkwood and Sterne, 2003]{Kirkwood2003}
Kirkwood, B.~R. and Sterne, J.~A. (2003).
\newblock {\em Essential Medical Statistics}.
\newblock Blackwell Publishing, 2nd edition.

\bibitem[Kuppens et~al., 2011]{Kuppens2011}
Kuppens, S., Holden, G., Barker, K., and Rosenberg, G. (2011).
\newblock A kappa-related decision: K, y, g, or ac1.
\newblock {\em Social Work Research}, 35.

\bibitem[Lienert, 1972]{Lienert1972}
Lienert, G. (1972).
\newblock Note on tests concerning the g index of agreement.
\newblock {\em Educational and Psychological Measurement}, 32.

\bibitem[Lu, 2010]{Lu2010}
Lu, Y. (2010).
\newblock A revised version of mcnemar's test for paired binary data.
\newblock {\em Communications in Statistics - Theory and Methods}, 39.

\bibitem[Lu et~al., 2017]{Lu2017}
Lu, Y., Wang, M., and Zhang, G. (2017).
\newblock A new revised version of mcnemar's test for paired binary data.
\newblock {\em Communications in Statistics - Theory and Methods}, 46.

\bibitem[Ludbrook, 2011]{Ludbrook2011}
Ludbrook, J. (2011).
\newblock Is there still a place for pearson's chi-squared test and fisher's
  exact test in surgical research?
\newblock {\em ANZ Journal of Surgery}, 81.

\bibitem[Manning et~al., 2008]{Manning2008}
Manning, C.~D., Raghavan, P., and Schutze, H. (2008).
\newblock {\em Introduction to Information Retrieval}.

\bibitem[Matthews, 1975]{Matthews1975}
Matthews, B.~W. (1975).
\newblock Comparison of the predicted and observed secondary structure of t4
  phage lysozyme.
\newblock {\em BBA - Protein Structure}, 405.

\bibitem[McNemar, 1947]{McNemar1947}
McNemar, Q. (1947).
\newblock Note on the sampling error of the difference between correlated
  proportions or percentages.
\newblock {\em Psychometrika}, 12:153--157.

\bibitem[Scott, 1955]{Scott1955}
Scott, W.~A. (1955).
\newblock Reliability of content analysis: The case of nominal scale coding.
\newblock {\em Public Opinion Quarterly}, 19.

\bibitem[Shankar and Bangdiwala, 2014]{Shankar2014}
Shankar, V. and Bangdiwala, S.~I. (2014).
\newblock Observer agreement paradoxes in 2x2 tables: Comparison of agreement
  measures.
\newblock {\em BMC Medical Research Methodology}, 14.

\bibitem[Shreiner, 1980]{Shreiner1980}
Shreiner, S.~C. (1980).
\newblock Agreement or association: Choosing a measure of reliability for
  nominal data in the 2 × 2 case - a comparison of phi, kappa, and g.
\newblock {\em Substance Use and Misuse}, 15.

\bibitem[Sim and Wright, 2005]{Sim2005}
Sim, J. and Wright, C.~C. (2005).
\newblock The kappa statistic in reliability studies: Use, interpretation, and
  sample size requirements.
\newblock {\em Physical Therapy}, 85.

\bibitem[Wongpakaran et~al., 2013]{Wongpakaran2013}
Wongpakaran, N., Wongpakaran, T., Wedding, D., and Gwet, K.~L. (2013).
\newblock A comparison of cohen's kappa and gwet's ac1 when calculating
  inter-rater reliability coefficients: A study conducted with personality
  disorder samples.
\newblock {\em BMC Medical Research Methodology}, 13.

\bibitem[Xie et~al., 2017]{Xie2017}
Xie, Z., Gadepalli, C., and Cheetham, B. M.~G. (2017).
\newblock Reformulation and generalisation of the cohen and fleiss kappas.
\newblock 3:16.

\bibitem[Yule, 1912]{Yule1912}
Yule, G.~U. (1912).
\newblock On the methods of measuring association between two attributes.
\newblock {\em Journal of the Royal Statistical Society}, 75.

\end{thebibliography}
\bibliographystyle{apalike}

\appendix

\section{R functions}\label{ap:Rfunctions}

All procedures applied along this text were implemented in R and are available to download from\newline\newline
https://sourceforge.net/projects/tables2x2.\newline\newline

Estimators were computed according the convention adopted on Table~\ref{tab:mathnotation}. In addition, implementations of several R packages were also incorporated in our experiments for two main reasons:
\begin{enumerate}
	\item to check if the point-estimate using the $abcd$-based formulas were correct, and
	\item to take advantage of inferential statistics already implemented in R packages.
\end{enumerate}

In some cases we could not locate any implementation in R packages, thus we applied bootstrapping to obtain, at least, a binary decision (rejection or non-rejection of the null hypothesis) based on confidence interval 95\%. Details are described in section ``\nameref{ap:G}".

In order to show the application of the following procedures, we adopted an example from the results of Bell and Kato-Katz examination performed on each of 315 stool specimens~\cite{Hoff1982}. \verb"Data" is a contingency table that can be incorporated in matrix \verb"m" with:

\begin{Verbatim}[frame=single, fontsize=\footnotesize]
Data <- ("
  BellxKK P   N
  P       184 54
  N       14  63
")
m <- as.matrix(read.table(textConnection(Data), 
                           header=TRUE, row.names=1))
print(m)                           
\end{Verbatim}
                           
Sometimes, according to the required input by some functions, an extensive presentation in data frame, \verb"dt", can be obtained with:                           
                           
\begin{Verbatim}[frame=single, fontsize=\footnotesize]
dt <- DescTools::Untable(m)
print(dt)                           
\end{Verbatim}

It follows the implementation of all estimator functions applied for the current work.

\subsection{Implementation of Cohen's kappa ($\kappa$) and corrected kappa ($\kappa_M$)}\label{ap:kappa}

The computation of Cohen's~$\kappa$~\cite{Cohen1960} and his proposed correction by maximum $\kappa$ were implemented by two functions:\mynobreakpar
\begin{itemize}
	\item The function \verb"agr2x2_kCohen" receives the parameters \verb"a", \verb"b", \verb"c", and \verb"d", or, alternatively, the matrix \verb"m" in place of the first parameter (in this case, internaly converting the matrix in \verb"a",~\verb"b",~\verb"c",~\verb"d"), to implement equation~\ref{eq:kappa_abcd}; it returns a matrix containing the essential computation: $\kappa$, $\kappa_M$ and $p$.  
	\item The auxiliary function \verb"agr2x2_maximum_kCohen" computes the maximum kappa~($\kappa_M$, equation~\ref{eq:kappamax}). It is called from \verb"agr2x2_kCohen" only when $\kappa>0$. 
\end{itemize} 

The inferential statistics is provided by \verb"epiR::epi.kappa", executed only if \verb"test=TRUE" to capture its $p$~value. 
Since Cohen's~$\kappa$ is a measure of agreement between observations, higher values of $\kappa$ lead to rejection of the null hypothesis~($H_0: \kappa=0$), thus providing evidence for agreement between observers or methods (generically named as raters along this text) when $\kappa$ is positive. Conversely, significantly negative values of $\kappa$ suggests disagreement between raters.

The implementation follows:\mynobreakpar
\begin{Verbatim}
# Cohen, J. (1960). A coefficient of agreement for nominal scales. 
# Educational and Psychological Measurement, 20(1).

# Auxiliary function to compute maximum Kappa
agr2x2_maximum_kCohen <- function(a,b,c,d)
{
  if (is.matrix(a))
  {
    m <- a
    a <- m[1,1]
    b <- m[1,2]
    c <- m[2,1]
    d <- m[2,2]
  }
  n <- a+b+c+d
  pc <- ((a+b)*(a+c)+(c+d)*(b+d))/(n^2)
  poM <- (min((a+c),(a+b))+min((b+d),(c+d)))/n
  kM <- (poM - pc)/(1-pc)
  
  return (kM)
}

# Cohen's kappa (k) and
# Correted(by maximum kappa) Cohen's kappa (kM)
agr2x2_kCohen <- function(a,b,c,d,test=FALSE)
{
  if (is.matrix(a))
  {
    m <- a
    a <- m[1,1]
    b <- m[1,2]
    c <- m[2,1]
    d <- m[2,2]
  }
  n <- a+b+c+d
  po <- (a+d)/n
  pc <- ((a+b)*(a+c)+(c+d)*(b+d))/(n^2)
  k <- (po-pc)/(1-pc)
  
  # computation of maximum kappa (when kappa is positive)
  kM <- k
  if (is.finite(k))
  {
    if(k>0)
    {
      kM <- agr2x2_maximum_kCohen(a,b,c,d)
    } 
  }
  
  # statistical test
  p.value <- NA
  if(test)
  {
    if (is.finite(k))
    {
      m <- matrix(data=c(a,b,c,d), nrow=2, ncol=2, byrow = TRUE)
      k.test <- epiR::epi.kappa(m,alternative="two.sided")
      p.value <- k.test$z$p.value
    }
  }
  
  m <- matrix(data=c(k,kM,p.value),nrow=1,ncol=3)
  colnames(m) <- c("k","kM","p")
  
  return(m)
}
\end{Verbatim}\label{fnk_k}

An example of use of this function, assessing Hoff et al.~\cite{Hoff1982} data is:\mynobreakpar
\begin{Verbatim}
# Agreement of Bell and Kato-Katz examination methods
Data <- ("
BellxKK P   N
P       184 54
N        14 63
")
m <- as.matrix(read.table(textConnection(Data),
                          header=TRUE, row.names=1))
print(m)
res <- agr2x2_kCohen(m,test=TRUE)
print(res)
\end{Verbatim}

\subsection{Implementation of Holley and Guilford's $G$}\label{ap:G}

Lienert~\cite{Lienert1972} proposed an inferential asymptotic statistical test for Holley \& Guilford~\cite{Holley1964} computing:
\begin{eqnarray}
\label{eq:u}
\begin{aligned}[b]
u = \frac{a+d-\frac{n}{2}}{\sqrt{\frac{n}{4}}}
\end{aligned}
\end{eqnarray}
For large samples ($n>30$) the statistic $(a+d)$ has distribution approximately normal with mean $\frac{n}{2}$ and variance $\frac{n}{4}$; consequently, $u$ has standard $z$ distribution, from which we can compute the correspondent two-sided $p$~values with R by \verb"2*(1-pnorm(abs(u)))" to the statistical decision under the null hypothesis, $H_0:G=0$.  

The function \verb"agr2x2_G" receives the same parameters and returns a matrix alike \verb"agr2x2_kCohen"~-~see~``\nameref{fnk_k}" for details~-~and implements the computation of $G$~(equation~\ref{eq:G}) and $u$~(equation~\ref{eq:u}):\mynobreakpar

\begin{Verbatim}
# G
# Holley & Guilford (1964), Lienert1972), Shreiner (1980)
agr2x2_G <- function(a,b,c,d,test=FALSE)
{
  if (is.matrix(a))
  {
    m <- a
    a <- m[1,1]
    b <- m[1,2]
    c <- m[2,1]
    d <- m[2,2]
  }
  n <- a+b+c+d
  G <- (a+d - (b+c)) / n # point estimate
  p.value <- NA
  if(test)
  {
    x <- a+d
    u <- (x-n/2)/sqrt(n/4)    
    p.value <- 1-pnorm(abs(u))
    p.value <- p.value*2
  }
  m <- matrix(data=c(G,p.value),nrow=1,ncol=2)
  colnames(m) <- c("G","p")
  return(m)
}

\end{Verbatim}

For example:
\mynobreakpar
\begin{Verbatim}
# Agreement of Bell and Kato-Katz examination methods
Data <- ("
BellxKK P   N
P       184 54
N        14 63
")
m <- as.matrix(read.table(textConnection(Data),
                          header=TRUE, row.names=1))
print(m)
res <- agr2x2_G(m,test=TRUE)
print(res)
\end{Verbatim}

\textbf{Bootstrapping}: In addition, since we also tested tables with $n \le 30$, it was implemented a version (called $SAC$) with inferential statistical decision by bootstrapping, which is a robust statistical method based on resamplings with replacements, independent of sample size and of variable distribution~\cite{Efron2007}. 

\small(\verb"agr2x2_SAC" receives parameters in similar fashion of \verb"agr2x2_f1Dice" to find a binary version of $p$~value~-~see ``\nameref{ap:Scott_pi}")\normalsize\mynobreakpar

The function \verb"agr2x2_SAC" is:\mynobreakpar
\begin{Verbatim}
# SAC

source("agr2x2_boot.table.R")

agr2x2_SAC <- function(a,b,c,d,B=FALSE,showboot=FALSE,plot=FALSE)
{
  if (is.matrix(a))
  {
    m <- a
    a <- m[1,1]
    b <- m[1,2]
    c <- m[2,1]
    d <- m[2,2]
  }
  sac <- (a+d - (b+c)) / (a+b+c+d) # point estimate
  sacLB <- sacUB <- pbin <- NA
  if(is.finite(sac))
  {
    if(B) # number of bootstraps
    {
      m <- matrix(data=c(a,b,c,d), nrow=2, ncol=2, byrow = TRUE)
      c_sac <- c() # bootstrap distribution of SACs
      if(showboot)
      {
        cat("\nbootstrapping\n")
        rotate <- c("/","-","\\","|")
        idxrotate <- 1
      }
      for (i in 1:B)
      {
        if(showboot)
        {
          cat(rotate[idxrotate],"\b",sep="")
          idxrotate <- idxrotate+1
          if(idxrotate>length(rotate)){idxrotate<-1}
          if(i%%B==0){cat(i,"resamplings\n")}
        }
        # variant 2x2 table
        m.new <- agr2x2_boot.table(m)
        a.new <- m.new[1,1]
        b.new <- m.new[1,2]
        c.new <- m.new[2,1]
        d.new <- m.new[2,2]
        # recursive call, point estimate only
        s <- agr2x2_SAC(a.new,b.new,c.new,d.new,B=FALSE)
        c_sac <- c(c_sac,as.numeric(s[[1]]))
      }
      c_sac <- c_sac[is.finite(c_sac)]
      if (length(c_sac)>0)
      {
        if(length(c_sac)>=2)
        {
          dsac <- density(c_sac, na.rm=TRUE)
          hdi <- HDInterval::hdi(dsac, credMass=0.95, allowSplit = TRUE)
          sacLB <- hdi[1]
          sacUB <- hdi[2]
          if(plot)
          {
            plot(dsac, xlab="SAC", ylab="Density",
                 main=paste0("Distribution of SAC\n",B," bootstraps"))
            yh <- max(dsac$y)/20; yi <- yh/4
            lines(c(sacLB,sacLB,sacLB,sacUB,sacUB,sacUB),
                  c(yh-yi,yh+yi,yh   ,yh   ,yh-yi,yh+yi))
            text((sacLB+sacUB)/2, yh+yi*2, "Prediction band 95%")
          }
          pbin <- 0
          if (sacLB<=0 & 0<=sacUB) {pbin <- 1}
        }
      }
    }
  }
  m <- matrix(data=c(sac,sacLB,sacUB,pbin),nrow=1,ncol=4)
  colnames(m) <- c("SAC","SACLB","SACUB","pbin")
  return(m)
}

\end{Verbatim}

Bootstrapping requires resampling with replacements. For that, 2x2~tables based on the table under investigation are generated, each one computing a value of $SAC$, implemented in function \verb"agr2x2_boot.table":\mynobreakpar
\begin{Verbatim}
agr2x2_boot.table <- function(m)
{
  a <- m[1,1]
  b <- m[1,2]
  c <- m[2,1]
  d <- m[2,2]
  n <- a+b+c+d
  # probabilities per cell
  probs <- c(a,b,c,d)/n
  acm_probs <- c()
  acm <- 0
  for (idx in 1:4)
  {
    acm <- acm+probs[idx]
    acm_probs <- c(acm_probs, acm)
  }
  # hypothetical new table  
  cells <- c(0,0,0,0)
  for (a in 1:n)
  {
    rnd=runif(1,min=0,max=1)  
    for (idx in 1:4)
    {
      if (rnd <= acm_probs[idx])
      {
        cells[idx] <- cells[idx]+1
        break
      }
    }
  }
  # proposed matrix
  return(matrix(data=cells, nrow=2, ncol=2, byrow = TRUE))
}
\end{Verbatim}

After thousands of repetitions (\verb"B" resamplings) the high density interval of the distribution of $SAC$ values (using \verb"HDInterval::hdi") can be checked. The prediction interval of $SAC$ is used to obtain a binary decision (``\nameref{ap:Scott_pi}" describes the rationale for \verb"pbin"): the non-rejection of this null hypothesis~(\verb"pbin=1") occurs when zero is inside the prediction band (it implies indefinition between raters), while its rejection~(\verb"pbin=0") is interpreted as non-null $SAC$ (which is evidence of agreement when $SAC>0$ or disagreement when $SAC<0$ between raters). For example, a bootstrapping with 100,000~resamplings (the function can also graphically show the 95\% prediction band) is obtained with:\mynobreakpar
\begin{Verbatim}
# Agreement of Bell and Kato-Katz examination methods
Data <- ("
BellxKK P   N
P       184 54
N        14 63
")
m <- as.matrix(read.table(textConnection(Data),
                          header=TRUE, row.names=1))
print(m)
res <- agr2x2_SAC(m,B=100000,showboot=TRUE,plot=TRUE)
print(res)
\end{Verbatim}

Under this procedure $SAC$ provided almost the same behavior of $G$~(Figure~\ref{fig:densityplots64}) with thousands of different tables~\verb"m", each one subjected to thousands of bootstraps. Therefore, the higher the absolute value of $G$ or $SAC$ the greater is the agreement (for positive numbers) or disagreement (for negative estimatives) between raters. 

\subsection{Implementation of Yule's Q}\label{ap:Q}

The function \verb"agr2x2_qYule" implements equation~\ref{eq:Q}.

\small(it receives the same parameters and returns a matrix alike \verb"agr2x2_kCohen"~-~see~``\nameref{fnk_k}").\normalsize 

When descriptive statistics is requested, it applies
\verb"exact2x2::fisher.exact". The null hypothesis of Yule's Q is the independence of rows and columns in a contingency table with fixed marginals. The rejection of the null hypothesis is interpreted as evidence of agreement~($Q>0$) or disagreement~($Q<0$) between raters:\mynobreakpar

\begin{Verbatim}
# Yule's Q
agr2x2_qYule <- function(a,b,c,d,test=FALSE)
{
  if (is.matrix(a))
  {
    m <- a
    a <- m[1,1]
    b <- m[1,2]
    c <- m[2,1]
    d <- m[2,2]
  }
  n <- a+b+c+d
  Q <- (a*d-b*c)/(a*d+b*c)
  p.value <- NA
  if (is.finite(Q))
  {
    if(test)
    {
      suppressMessages(library(exact2x2))
      m <- matrix(data=c(a,b,c,d), nrow=2, ncol=2, byrow = TRUE)
      f.test <- NA
      f.test <- exact2x2::fisher.exact(m)
      p.value <- f.test$p.value
    }
  }
  m <- matrix(data=c(Q,p.value),nrow=1,ncol=2)
  colnames(m) <- c("Q","p.value")
  return(m)
}
\end{Verbatim}\label{fnc_Q}

For example:\mynobreakpar
\begin{Verbatim}
# Agreement of Bell and Kato-Katz examination methods
Data <- ("
BellxKK P   N
P       184 54
N        14 63
")
m <- as.matrix(read.table(textConnection(Data),
                          header=TRUE, row.names=1))
print(m)
res <- agr2x2_qYule(m,test=TRUE)
print(res)
\end{Verbatim}

\subsection{Implementation of Yule's Y}\label{ap:Y}

Similar to Yule's~$Q$, this estimator was implemented by \verb"agr2x2_yYule"~(equation~\ref{eq:Y}). Since we did not locate any implementation of this estimator nor any suitable inferential statistical test, it was implemented by bootstrapping (it also depends on \verb"agr2x2_boot.table", which is described in ``~\nameref{ap:G}"~-~see $SAC$). 

The function \verb"agr2x2_yYule" was implemented as:\mynobreakpar
\begin{Verbatim}
# Yule's Y

source("agr2x2_boot.table.R")

agr2x2_yYule <- function(a,b,c,d,B=FALSE,showboot=FALSE,plot=FALSE)
{
  if (is.matrix(a))
  {
    m <- a
    a <- m[1,1]
    b <- m[1,2]
    c <- m[2,1]
    d <- m[2,2]
  }
  n <- a+b+c+d
  Y <- (sqrt(a*d)-sqrt(b*c))/
       (sqrt(a*d)+sqrt(b*c))
  YLB <- YUB <- pbin <- NA
  if(is.finite(Y))
  {
    if(B) # number of bootstraps
    {
      m <- matrix(data=c(a,b,c,d), nrow=2, ncol=2, byrow = TRUE)
      c_Y <- c() # bootstrap distribution of F1s (adjusted)
      if(showboot)
      {
        cat("\nbootstrapping\n")
        rotate <- c("/","-","\\","|")
        idxrotate <- 1
      }
      for (i in 1:B)
      {
        if(showboot)
        {
          cat(rotate[idxrotate],"\b",sep="")
          idxrotate <- idxrotate+1
          if(idxrotate>length(rotate)){idxrotate<-1}
          if(i%%B==0){cat(i,"resamplings\n")}
        }
        # variant 2x2 table
        m.new <- agr2x2_boot.table(m)
        a.new <- m.new[1,1]
        b.new <- m.new[1,2]
        c.new <- m.new[2,1]
        d.new <- m.new[2,2]
        # recursive call, point estimate only, no test, no graph
        s <- (sqrt(a.new*d.new)-sqrt(b.new*c.new))/
          (sqrt(a.new*d.new)+sqrt(b.new*c.new))
        c_Y <- c(c_Y,s)
      }
      c_Y <- c_Y[is.finite(c_Y)]
      if (length(c_Y)>0)
      {
        if(length(c_Y)>=2)
        {
          dY <- density(c_Y, na.rm=TRUE)
          hdi <- HDInterval::hdi(dY, credMass=0.95, allowSplit = TRUE)
          YLB <- hdi[1]
          YUB <- hdi[2]
          if(plot)
          {
            plot(dY, xlab="Yule's Y", ylab="Density",
                 main=paste0("Distribution of Yule's Y\n",B," bootstraps"))
            yh <- max(dY$y)/20; yi <- yh/4
            lines(c(YLB,YLB,YLB,YUB,YUB,YUB),
                  c(yh-yi,yh+yi,yh   ,yh   ,yh-yi,yh+yi))
            text((YLB+YUB)/2, yh+yi*2, "Prediction band 95%")
          }
          pbin <- 0
          if (YLB<=0 & 0<=YUB) {pbin <- 1}
        }
      }
    }
  }
  m <- matrix(data=c(Y,YLB,YUB,pbin),nrow=1,ncol=4)
  colnames(m) <- c("Y","Y.LB","Y.LB","pbin")
  return(m)
}
\end{Verbatim}

For example:
\begin{Verbatim}
# Agreement of Bell and Kato-Katz examination methods
Data <- ("
BellxKK P   N
P       184 54
N        14 63
")
m <- as.matrix(read.table(textConnection(Data),
                          header=TRUE, row.names=1))
print(m)
res <- agr2x2_yYule(m,B=100000,showboot=TRUE,plot=TRUE)
print(res)
\end{Verbatim}

\subsection{Implementation of Pearson's $r$}\label{ap:r}

The function \verb"agr2x2_rPearson" implements equation~\ref{eq:r}, receiving the same parameters and returning a matrix alike \verb"agr2x2_kCohen"~(see~``\nameref{fnk_k}"). When descriptive statistics is requested, it applies the function \verb"cor.test". Observe that \verb"cor.test" requires two numeric vectors to compute and test Pearson's correlation, thus it has to be preceded by \verb"DescTools::Untable". The null hypothesis is $H_0: \rho=0$, i.e., absence of correlation. The rejection of this null hypothesis is interpreted as evidence of agreement~($r>0$) or disagreement~($r<0$) between raters. 

Implementation of \verb"agr2x2_rPearson" is:\mynobreakpar
\begin{Verbatim}
# Pearson's r
agr2x2_rPearson <- function(a,b,c,d,test=FALSE)
{
  if (is.matrix(a))
  {
    m <- a
    a <- m[1,1]
    b <- m[1,2]
    c <- m[2,1]
    d <- m[2,2]
  }
  n <- a+b+c+d
  r <- (a*d-b*c)/sqrt((a+b)*(a+c)*(b+d)*(c+d))
  p.value <- NA
  if(test)
  {
    m <- matrix(data=c(a,b,c,d), nrow=2, ncol=2, byrow = TRUE)
    dt <- DescTools::Untable(m)
    try(
      ct <- cor.test(as.numeric(dt$Var1),as.numeric(dt$Var2)),
      silent=TRUE
    )
    try(
      p.value <- ct$p.value,
      silent=TRUE
    )
  }
  m <- matrix(data=c(r,p.value),nrow=1,ncol=2)
  colnames(m) <- c("r","p.value")
  return(m)
}
\end{Verbatim}\label{fnc_r}

For example:\mynobreakpar
\begin{Verbatim}
# Agreement of Bell and Kato-Katz examination methods
Data <- ("
BellxKK P   N
P       184 54
N        14 63
")
m <- as.matrix(read.table(textConnection(Data),
                          header=TRUE, row.names=1))
print(m)
res <- agr2x2_rPearson(m,test=TRUE)
print(res)
\end{Verbatim}

We emphasize that Pearson's~$r$ provides the same results of Matthews' correlation coefficient, Cramér's~$V$, Pearson's~$\phi$ (from which Pearson's contingent coefficient is function, equation~\ref{eq:PCC}), Yule's~$\phi$, Spearman's~$\rho$, and Kendall's~$\tau$, reason for the omission of separed analysis of all these estimators in 2x2~tables (see~``\nameref{tit:r}").

\subsection{Implementation of McNemar's $\chi^2$}\label{ap:McNemar}

The function \verb"agr2x2_mnMcNemar" implements equation~\ref{eq:MN}, which corresponds to the Normalized version of McNemar's~$\chi^2$. It receives the same parameters and returning a matrix alike \verb"agr2x2_kCohen"~(see~``\nameref{fnk_k}"). When descriptive statistics is requested, it applies bootstrapping (described in ``~\nameref{ap:G}"~-~see $SAC$). 

Unlike other tests, the null hipothesis of McNemar's test is a test of change, therefore the rejection of the null hypothesis is taken as change and, therefore, not necessarily evidence of agreement or disagreement between raters~(see section~``\nameref{sec:discussion}"). 

It was implemented by:\mynobreakpar
\begin{Verbatim}
# McNemar's X^2 normalized
# MN

source("agr2x2_boot.table.R")

agr2x2_mnMcNemar <- function(a,b,c,d,B=FALSE,showboot=FALSE,plot=FALSE)
{
  if (is.matrix(a))
  {
    m <- a
    a <- m[1,1]
    b <- m[1,2]
    c <- m[2,1]
    d <- m[2,2]
  }
  MN <- abs(b-c)/(b+c)
  MNLB <- MNUB <- pbin <- NA
  if(is.finite(MN))
  {
    if(B) # number of bootstraps
    {
      m <- matrix(data=c(a,b,c,d), nrow=2, ncol=2, byrow = TRUE)
      c_MN <- c() # bootstrap distribution of MNs
      if(showboot)
      {
        cat("\nbootstrapping\n")
        segsize <- floor(B/10)/2
      }
      for (i in 1:B)
      {
        if(showboot)
        {
          if(i%%segsize==0){cat(".")}
          if(i%%B==0){cat(i,"resamplings\n")}
        }
        # variant 2x2 table
        m.new <- agr2x2_boot.table(m)
        a.new <- m.new[1,1]
        b.new <- m.new[1,2]
        c.new <- m.new[2,1]
        d.new <- m.new[2,2]
        s <- abs(b.new-c.new)/(b.new+c.new)
        c_MN <- c(c_MN,s)
      }
      c_MN <- c_MN[is.finite(c_MN)]
      if (length(c_MN)>0)
      {
        if(length(c_MN)>=2)
        {
          dMN <- density(c_MN, na.rm=TRUE)
          hdi <- HDInterval::hdi(dMN, credMass=0.95, allowSplit = TRUE)
          MNLB <- hdi[1]
          MNUB <- hdi[2]
          if(plot)
          {
            plot(dMN, xlab="Normalized McNemar's X^2 (MN)",
                 ylab="Density",
                 main=paste0("Distribution of MN\n",B," bootstraps"))
            yh <- max(dMN$y)/20; yi <- yh/4
            lines(c(MNLB,MNLB,MNLB,MNUB,MNUB,MNUB),
                  c(yh-yi,yh+yi,yh   ,yh   ,yh-yi,yh+yi))
            text((MNLB+MNUB)/2, yh+yi*2, "Prediction band 95%")
          }
          pbin <- 0
          if (MNLB<=0 & 0<=MNUB) {pbin <- 1}
        }
      }
    }
  }
  m <- matrix(data=c(MN,MNLB,MNUB,pbin),nrow=1,ncol=4)
  colnames(m) <- c("MN","MN.LB","MN.UB","pbin")
  return(m)
}
\end{Verbatim}\label{fnc_MN}

With the example of Bell and Kato-Katz methods, this function can be called by:\mynobreakpar
\begin{Verbatim}
# Agreement of Bell and Kato-Katz examination methods
Data <- ("
BellxKK P   N
P       184 54
N        14 63
")
m <- as.matrix(read.table(textConnection(Data),
                          header=TRUE, row.names=1))
print(m)
res <- agr2x2_mnMcNemar(m,B=100000,showboot=TRUE,plot=TRUE)
print(res)
\end{Verbatim}

Although not analized along the main text, the Traditional McNemar's~$\chi^2$ was also implemented in \verb"agr2x2_chi2mnMcNemar". When the inferential test is required, it is performed by \verb"exact2x2::mcnemar.exact" computing the estimative of probability-ratio and confidence interval given by $\frac{b}{c}$; the null hypothesis is rejected when the unitary value is not included in its confidence interval 95\%. It was implemented by:\mynobreakpar
\begin{Verbatim}
# McNemar's X2
agr2x2_chi2mnMcNemar <- function(a,b,c,d,test=FALSE)
{
  if (is.matrix(a))
  {
    m <- a
    a <- m[1,1]
    b <- m[1,2]
    c <- m[2,1]
    d <- m[2,2]
  }
  n <- a+b+c+d

  Chi2MN <- ((b-c)^2)/(b+c)
  p.value <- OR <- OR.LB <- OR.UB <- NA
  if(is.finite(Chi2MN))
  {
    if(test)
    {
      suppressMessages(library(exact2x2))
      m <- matrix(data=c(a,b,c,d), nrow=2, ncol=2, byrow = TRUE)
      Chi2.test <- exact2x2::mcnemar.exact(m)
      p.value <- Chi2.test$p.value
      OR <- as.numeric(Chi2.test$estimate)
      OR.LB <- as.numeric(Chi2.test$conf.int[1])
      OR.UB <- as.numeric(Chi2.test$conf.int[2])
    }
  }
  m <- matrix(data=c(Chi2MN,OR,OR.LB,OR.UB,p.value),nrow=1,ncol=5)
  colnames(m) <- c("Chi2MN","b/c","b/c.LB","b/c.UB","p")
  return(m)
}
\end{Verbatim}\label{fnc_chi2MN}

Both, Normalized and Traditional McNemar's $\chi^2$ only provide positive values computed from the off-diagonal, thus they cannot distinguish situations of agreement of disagreement between raters. Attempts to improve McNemar's $\chi^2$ are also registered~\cite{Lu2010,Lu2017}, which were implemented by:\mynobreakpar
\begin{Verbatim}
# McNemar's X2, revised 2010
# Lu, Y.: A revised version of mcnemar’s test for paired binary data.
# Communications in Statistics -  Theory and Methods 39 (2010).
# DOI 10.1080/03610920903289218

source("agr2x2_boot.table.R")

agr2x2_mnMcNemar2010 <- function(a,b,c,d,B=FALSE,
                                            showboot=FALSE,plot=FALSE)
{
  if (is.matrix(a))
  {
    m <- a
    a <- m[1,1]
    b <- m[1,2]
    c <- m[2,1]
    d <- m[2,2]
  }
  n <- a+b+c+d
  MN2010 <- ((b-c)^2)/((b+c)*(1+((a+b)/(a+b+c+d))))
  MN2010LB <- MN2010UB <- pbin2010 <- NA
  if(is.finite(MN2010))
  {
    if(B) # number of bootstraps
    {
      m <- matrix(data=c(a,b,c,d), nrow=2, ncol=2, byrow = TRUE)
      c_MN2010 <- c() # bootstrap distribution of MN2010
      if(showboot)
      {
        cat("\nbootstrapping\n")
        segsize <- floor(B/10)/2
      }
      for (i in 1:B)
      {
        if(showboot)
        {
          if(i%%segsize==0){cat(".")}
          if(i%%B==0){cat(i,"resamplings\n")}
        }
        # variant 2x2 table
        m.new <- agr2x2_boot.table(m)
        a.new <- m.new[1,1]
        b.new <- m.new[1,2]
        c.new <- m.new[2,1]
        d.new <- m.new[2,2]
        # point estimate only
        s2010 <- ((b.new-c.new)^2)/((b.new+c.new)*
                 (1+((a.new+b.new)/(a.new+b.new+c.new+d.new) )))
        c_MN2010 <- c(c_MN2010,s2010)
      }
      c_MN2010 <- c_MN2010[is.finite(c_MN2010)]
      if (length(c_MN2010)>0)
      {
        if(length(c_MN2010)>2)
        {
          d2010 <- density(c_MN2010, na.rm=TRUE)
          hdi <- HDInterval::hdi(d2010, credMass=0.95,
                                 allowSplit = FALSE)
          MN2010LB <- hdi[1]
          MN2010UB <- hdi[2]
          pbin2010 <- 0
          if (MN2010LB<=0 & 0<=MN2010UB) {pbin2010 <- 1}
        }
        if(plot)
        {
          if(length(c_MN2010)>2)
          {
            plot(d2010, xlab="McNemar's X², revised by Yu (2010)",
                 ylab="Density",
                 main=paste0("Distribution of MN\n",B," bootstraps"))
            yh <- max(d2010$y)/20; yi <- yh/4
            lines(c(MN2010LB,MN2010LB,MN2010LB,
                    MN2010UB,MN2010UB,MN2010UB),
                  c(yh-yi,yh+yi,yh   ,
                    yh   ,yh-yi,yh+yi))
            text((MN2010LB+MN2010UB)/2, yh+yi*2,
                 "Prediction band 95%")
          }
        }
      }
    }
  }
  m <- matrix(data=c(MN2010,MN2010LB,MN2010UB,pbin2010),nrow=1,ncol=4)
  colnames(m) <- c("MN2010","MN2010LB","MN2010UB","pbin2010")
  return(m)
}
\end{Verbatim}\label{fnc_MN2010}
\begin{Verbatim}
# McNemar's X2
# Lu, Y., Wang, M., Zhang, G.: A new revised version of mcnemar’s test
# for paired binary data. Communications in Statistics -
# Theory and Methods 46 (2017).
# DOI 10.1080/03610926.2016.1228962

source("agr2x2_boot.table.R")

agr2x2_mnMcNemar2017 <- function(a,b,c,d,B=FALSE,
                                     showboot=FALSE,plot=FALSE)
{
  if (is.matrix(a))
  {
    m <- a
    a <- m[1,1]
    b <- m[1,2]
    c <- m[2,1]
    d <- m[2,2]
  }
  n <- a+b+c+d
  MN2017 <- (((a+b+c+d)*(b-c)^2)/((b+c+2*a)*(b+c+2*d)))
  MN2017LB <- MN2017UB <- pbin2017 <- NA
  if(is.finite(MN2017))
  {
    if(B) # number of bootstraps
    {
      m <- matrix(data=c(a,b,c,d), nrow=2, ncol=2, byrow = TRUE)
      c_MN2017 <- c() # bootstrap distribution of MN2017
      if(showboot)
      {
        cat("\nbootstrapping\n")
        segsize <- floor(B/10)/2
      }
      for (i in 1:B)
      {
        if(showboot)
        {
          if(i%%segsize==0){cat(".")}
          if(i%%B==0){cat(i,"resamplings\n")}
        }
        # variant 2x2 table
        m.new <- agr2x2_boot.table(m)
        a.new <- m.new[1,1]
        b.new <- m.new[1,2]
        c.new <- m.new[2,1]
        d.new <- m.new[2,2]
        # point estimate only
        s2017 <- (((a.new+b.new+c.new+d.new)*(b.new-c.new)^2)/
                    ((b.new+c.new+2*a.new)*(b.new+c.new+2*d.new)))
        # s2017 <- (n*abs(b.new-c.new))/
        #          ((b.new+c.new+2*a.new)*(b.new+c.new+2*d.new))
        c_MN2017 <- c(c_MN2017,s2017)
      }
      c_MN2017 <- c_MN2017[is.finite(c_MN2017)]
      if (length(c_MN2017)>0)
      {
        if(length(c_MN2017)>2)
        {
          d2017 <- density(c_MN2017, na.rm=TRUE)
          hdi <- HDInterval::hdi(d2017, credMass=0.95,
                                 allowSplit = FALSE)
          MN2017LB <- hdi[1]
          MN2017UB <- hdi[2]
          pbin2017 <- 0
          if (MN2017LB<=0 & 0<=MN2017UB) {pbin2017 <- 1}
        }
        if(plot)
        {
          if(length(c_MN2017)>2)
          {
            plot(d2017, xlab="McNemar's X², revised by Yu (2017)",
                 ylab="Density",
                 main=paste0("Distribution of MN\n",B," bootstraps"))
            yh <- max(d2017$y)/20; yi <- yh/4
            lines(c(MN2017LB,MN2017LB,MN2017LB,
                    MN2017UB,MN2017UB,MN2017UB),
                  c(yh-yi,yh+yi,yh   ,
                    yh   ,yh-yi,yh+yi))
            text((MN2017LB+MN2017UB)/2, yh+yi*2,
                 "Prediction band 95%")
          }
        }
      }
    }
  }
  m <- matrix(data=c(MN2017,MN2017LB,MN2017UB,pbin2017),nrow=1,ncol=4)
  colnames(m) <- c("MN2017","MN2017LB","MN2017UB","pbin2017")
  return(m)
}
\end{Verbatim}\label{fnc_MN2017}

These three versions of the Traditional McNemar's $\chi^2$ can be called by:\mynobreakpar
\begin{Verbatim}
# Agreement of Bell and Kato-Katz examination methods
Data <- ("
BellxKK P   N
P       184 54
N        14 63
")
m <- as.matrix(read.table(textConnection(Data),
                          header=TRUE, row.names=1))
print(m)
res <- agr2x2_chi2mnMcNemar(m,test=TRUE)
print(res)
res <- agr2x2_mnMcNemar2010(m,B=100000,
                                       showboot=TRUE,plot=TRUE)
print(res)
res <- agr2x2_mnMcNemar2017(m,B=100000,
                                       showboot=TRUE,plot=TRUE)
print(res)

\end{Verbatim}

All alternatives were tested with the scripts available along this supplemental material. None of them are fit for the general cases of agreement as discussed in the main text and were not included in the main analysis.

\subsection{Implementation of Scott's $\pi$}\label{ap:Scott_pi}

The function \verb"agr2x2_piScott" implements equation~\ref{eq:pi}, receiving the same parameters and returning a matrix alike \verb"agr2x2_kCohen"~(see~``\nameref{fnk_k}"). When descriptive statistics is requested, it applies the function \verb"rel::spi" (it requires \verb"DescTools::Untable", see~``\nameref{ap:r}"). 

Unfortunately, \verb"rel::spi" provides only a confidence interval 95\% of $\pi$ without $p$~value. The null hipothesis is $H_0: \pi=0$, i.e., absence of agreement or disagreement. Lacking a $p$~value, our proposition is a binary decision, a ``binary $p$~value", assuming the non-rejection of this null hypothesis when \verb"pbin=1", implying indefinition between raters (neutrality). Since Scott's~$\pi$ is an index of agreement between raters, when the null hypothesis is rejected~(\verb"pbin=0"), it implies that the higher the absolute value the greater is the agreement~(if $\pi>0$) or the disagreement~(if $\pi<0$) between raters. 

It is implemented by\mynobreakpar
\begin{Verbatim}
# Scott's pi
agr2x2_piScott <- function(a,b,c,d,test=FALSE)
{
  if (is.matrix(a))
  {
    m <- a
    a <- m[1,1]
    b <- m[1,2]
    c <- m[2,1]
    d <- m[2,2]
  }
  n <- a+b+c+d
  po <- (a+d)/n
  pc <- (((a+c+a+b)/(2*n))^2)+(((c+d+b+d)/(2*n))^2)
  spi <- (po-pc)/(1-pc)

  pbin <- pi.lb <- pi.ub <- NA
  if(test)
  {
    if(is.finite(spi))
    {
      m <- matrix(data=c(a,b,c,d), nrow=2, ncol=2, byrow = TRUE)
      dt <- DescTools::Untable(m)
      # pi de Scott (kappa de Fleiss)
      pi.Scott <- NULL
      try(
        pi.Scott <- rel::spi(data=dt),
          silent=TRUE
      )
      if(length(pi.Scott[1])>0)
      {
        if (is.finite(pi.Scott$lb) & is.finite(pi.Scott$ub))
        {
          pi.lb <- pi.Scott$lb
          pi.ub <- pi.Scott$ub
          pbin <- 0
          if (pi.Scott$lb<=0 & 0<=pi.Scott$ub)
          {
            pbin <- 1
          }
        } else
        {
          pbin <- NaN
        }
      }
    }
  }
  m <- matrix(data=c(spi,pi.lb,pi.ub,pbin),nrow=1,ncol=4)
  colnames(m) <- c("pi","piLB","piUB","p_bin")
  return(m)
}
\end{Verbatim}

For example:\mynobreakpar
\begin{Verbatim}
# Agreement of Bell and Kato-Katz examination methods
Data <- ("
BellxKK P   N
P       184 54
N        14 63
")
m <- as.matrix(read.table(textConnection(Data),
                          header=TRUE, row.names=1))
print(m)
res <- agr2x2_piScott(m,test=TRUE)
print(res)
\end{Verbatim}

Scott's~$\pi$ provides the same results of Fleiss'~$\kappa$ in 2x2~tables, reason for the omission of the latter in our computations~(see~``\nameref{tit:scott_pi}").

\subsection{Implementation of Dice's $F1$ and Adjusted $F1_{adj}$}\label{ap:F1}

The function \verb"agr2x2_f1Dice" receives the same parameters and returning a matrix alike \verb"agr2x2_kCohen"~(see~``\nameref{fnk_k}").

Since it is a measure of agreement providing only positive numbers (ranging from 0 to 1), it is assumed that the higher the value of $F1$, the greater is the agreement between raters (0 is disagreement and 0.5 is neutrality). We proposed rescalling of $F1$~(Equation~\ref{eq:F1}) to $F1_{adj}$~(Equation~\ref{eq:F1adj}). 

To our knowledge, there is no inferential test implemented to date in R packages. Therefore, we countour this problem with bootstrapping (see bootstrapping on~``\nameref{ap:G}" for details). 

Equations~\ref{eq:F1} and \ref{eq:F1adj} and inferential tests were implemented as:\mynobreakpar
\begin{Verbatim}
# Dice's F1
# Adjusted F1

source("agr2x2_boot.table.R")

agr2x2_f1Dice <- function(a,b,c,d,B=FALSE,showboot=FALSE,plot=FALSE)
{
  if (is.matrix(a))
  {
    m <- a
    a <- m[1,1]
    b <- m[1,2]
    c <- m[2,1]
    d <- m[2,2]
  }
  f1 <- (2*a)/(2*a + b + c)
  f1.adj <- (2*a - (b+c)) / (2*a + (b+c))
  f1LB <- f1UB <- pbin <- NA
  if(is.finite(f1))
  {
    if(B) # number of bootstraps
    {
      m <- matrix(data=c(a,b,c,d), nrow=2, ncol=2, byrow = TRUE)
      c_f1 <- c() # bootstrap distribution of F1s (adjusted)
      if(showboot)
      {
        cat("\nbootstrapping\n")
        rotate <- c("/","-","\\","|")
        idxrotate <- 1
      }
      for (i in 1:B)
      {
        if(showboot)
        {
          cat(rotate[idxrotate],"\b",sep="")
          idxrotate <- idxrotate+1
          if(idxrotate>length(rotate)){idxrotate<-1}
          if(i%%B==0){cat(i,"resamplings\n")}
        }
        # variant 2x2 table
        m.new <- agr2x2_boot.table(m)
        a.new <- m.new[1,1]
        b.new <- m.new[1,2]
        c.new <- m.new[2,1]
        d.new <- m.new[2,2]
        # recursive call, point estimate only, no test, no graph
        s <- agr2x2_f1Dice(a.new,b.new,c.new,d.new)
        c_f1 <- c(c_f1,as.numeric(s[[1]]))
      }
      c_f1 <- c_f1[is.finite(c_f1)]
      if (length(c_f1)>0)
      {
        if(length(c_f1)>=2)
        {
          df1 <- density(c_f1, na.rm=TRUE)
          hdi <- HDInterval::hdi(df1, credMass=0.95,
                                 allowSplit = FALSE)
          f1LB <- hdi[1]
          f1UB <- hdi[2]
          if(plot)
          {
            plot(df1, xlab="Adjusted Dice's F1", ylab="Density",
                 main=paste0("Distribution of Dice's F1\n",
                             B," bootstraps"))
            yh <- max(df1$y)/20; yi <- yh/4
            lines(c(f1LB,f1LB,f1LB,f1UB,f1UB,f1UB),
                  c(yh-yi,yh+yi,yh   ,yh   ,yh-yi,yh+yi))
            text((f1LB+f1UB)/2, yh+yi*2, "Prediction band 95%")
          }
          pbin <- 0
          if (f1LB<=0 & 0<=f1UB) {pbin <- 1}
        }
      }
    }
  }
  m <- matrix(data=c(f1,f1.adj,f1LB,f1UB,pbin),nrow=1,ncol=5)
  colnames(m) <- c("F1","F1.adj","F1.adjLB","F1.adjUB","pbin")
  return(m)
}
\end{Verbatim}

The inferential statistics is performed by bootstrapping with the parameter \verb"B". The parameter \verb"showboot" is useful for long bootstrappings (showing the user that the computer is not freezed), and the distribution of $F1$ can be shown with \verb"plot=TRUE".  For example, a bootstrapping with 100,000~resamplings (the function can also graphically show the 95\% prediction band) is obtained with:
\begin{Verbatim}
# Agreement of Bell and Kato-Katz examination methods
Data <- ("
BellxKK P   N
P       184 54
N        14 63
")
m <- as.matrix(read.table(textConnection(Data),
                          header=TRUE, row.names=1))
print(m)
res <- agr2x2_f1Dice(m,B=100000,showboot=TRUE,plot=TRUE)
print(res)
\end{Verbatim}
The decision is binary~(see \verb"pbin" in ``\nameref{ap:Scott_pi}"), with rejection of the null hypothesis, $H_0: F1_{adj}=0$, when zero is out of the prediction band (i.e, \verb"pbin = 0"), thus providing evidence of agreement or disagreement between raters, depending on the observation of the 2x2~contingency table. 

\subsection{Implementation of Shankar and Bangdiwala's $B$}\label{ap:B}

This coefficient provides only positive numbers (ranging from 0 to 1). It is assumed that the higher the value of $B$, the greater is the agreement between raters (0 is disagreement and 0.5 is neutrality), depending on the observation of the 2x2~contingency table.

The implementation of equation~\ref{eq:B} is:\mynobreakpar
\begin{Verbatim}
# Agreement Plot

source("agr2x2_boot.table.R")

agr2x2_bShankar <- function(a,b,c,d,B=FALSE,
                            showboot=FALSE,plot=FALSE)
{
  if (is.matrix(a))
  {
    m <- a
    a <- m[1,1]
    b <- m[1,2]
    c <- m[2,1]
    d <- m[2,2]
  }
  SB <- (a^2+d^2)/((a+c)*(a+b)+(b+d)*(c+d))
  pbin <- NA
  SB_LB <- SB_UB <- pbin <- NA
  if(is.finite(SB))
  {
    if(B) # number of bootstraps
    {
      m <- matrix(data=c(a,b,c,d), nrow=2, ncol=2, byrow = TRUE)
      c_SB <- c() # bootstrap distribution of SBs
      if(showboot)
      {
        cat("\nbootstrapping\n")
        rotate <- c("/","-","\\","|")
        idxrotate <- 1
      }
      for (i in 1:B)
      {
        if(showboot)
        {
          cat(rotate[idxrotate],"\b",sep="")
          idxrotate <- idxrotate+1
          if(idxrotate>length(rotate)){idxrotate<-1}
          if(i%%B==0){cat(i,"resamplings\n")}
        }
        # variant 2x2 table
        m.new <- agr2x2_boot.table(m)
        a.new <- m.new[1,1]
        b.new <- m.new[1,2]
        c.new <- m.new[2,1]
        d.new <- m.new[2,2]
        # recursive call, point estimate only
        s <- (a.new^2+d.new^2)/((a.new+c.new)*(a.new+b.new)+
                                (b.new+d.new)*(c.new+d.new))
        c_SB <- c(c_SB,s)
      }
      c_SB <- c_SB[is.finite(c_SB)]
      if (length(c_SB)>0)
      {
        if(length(c_SB)>=2)
        {
          dsb <- density(c_SB, na.rm=TRUE)
          hdi <- HDInterval::hdi(dsb, credMass=0.95,
                                 allowSplit = FALSE)
          SB_LB <- hdi[1]
          SB_UB <- hdi[2]
          if(plot)
          {
            plot(dsb,
                 xlab="Bangdiwala and Shankar's coefficient",
                 ylab="Density",
                 main=paste0("Distribution of SB\n",B," bootstraps"))
            yh <- max(dsb$y)/20; yi <- yh/4
            lines(c(SB_LB,SB_LB,SB_LB,SB_UB,SB_UB,SB_UB),
                  c(yh-yi,yh+yi,yh   ,yh   ,yh-yi,yh+yi))
            text((SB_LB+SB_UB)/2, yh+yi*2, "Prediction band 95%")
          }
          pbin <- 0
          if (SB_LB<=0 & 0<=SB_UB) {pbin <- 1}
        }
      }
    }
  }
  m <- matrix(data=c(SB,SB_LB,SB_UB,pbin),nrow=1,ncol=4)
  colnames(m) <- c("SB","SB_LB","SB_UB","pbin")
  return(m)
}

\end{Verbatim}

The inferential statistical decision depends on bootstrapping (see bootstrapping on~``\nameref{ap:G}" for details); for example:\mynobreakpar
\begin{Verbatim}
# Agreement of Bell and Kato-Katz examination methods
Data <- ("
BellxKK P   N
P       184 54
N        14 63
")
m <- as.matrix(read.table(textConnection(Data),
                          header=TRUE, row.names=1))
print(m)
res <- agr2x2_bShankar(m,B=100000,showboot=TRUE,plot=TRUE)
print(res)
\end{Verbatim}

Our proposition for rescalling $B$ (equation~\ref{eq:B2}) is similar to the procedure of ``\nameref{ap:F1}". It was implemented as:
\mynobreakpar
\begin{Verbatim}
# Agreement Plot

source("agr2x2_boot.table.R")

agr2x2_badjShankar <- function(a,b,c,d,B=FALSE,
                               showboot=FALSE,plot=FALSE)
{
  if (is.matrix(a))
  {
    m <- a
    a <- m[1,1]
    b <- m[1,2]
    c <- m[2,1]
    d <- m[2,2]
  }
  SB <- (a^2+d^2)/((a+c)*(a+b)+(b+d)*(c+d))
  SB <- (SB*2)-1
  pbin <- NA
  SB_LB <- SB_UB <- pbin <- NA
  if(is.finite(SB))
  {
    if(B) # number of bootstraps
    {
      m <- matrix(data=c(a,b,c,d), nrow=2, ncol=2, byrow = TRUE)
      c_SB <- c() # bootstrap distribution of SBs
      if(showboot)
      {
        rotate <- c("/","-","\\","|")
        idxrotate <- 1
      }
      for (i in 1:B)
      {
        if(showboot)
        {
          cat(rotate[idxrotate],"\b",sep="")
          idxrotate <- idxrotate+1
          if(idxrotate>length(rotate)){idxrotate<-1}
          if(i%%B==0){cat(i,"resamplings\n")}
        }
        # variant 2x2 table
        m.new <- agr2x2_boot.table(m)
        a.new <- m.new[1,1]
        b.new <- m.new[1,2]
        c.new <- m.new[2,1]
        d.new <- m.new[2,2]
        # recursive call, point estimate only
        s <- (a.new^2+d.new^2)/((a.new+c.new)*(a.new+b.new)+
                                (b.new+d.new)*(c.new+d.new))
        s <- (s*2)-1
        c_SB <- c(c_SB,s)
      }
      c_SB <- c_SB[is.finite(c_SB)]
      if (length(c_SB)>0)
      {
        if(length(c_SB)>=2)
        {
          dsb <- density(c_SB, na.rm=TRUE)
          hdi <- HDInterval::hdi(dsb, credMass=0.95,
                                 allowSplit = FALSE)
          SB_LB <- hdi[1]
          SB_UB <- hdi[2]
          if(plot)
          {
            plot(dsb,
                 xlab="Bangdiwala and Shankar's coefficient",
                 ylab="Density",
                 main=paste0("Distribution of SB\n",B," bootstraps"))
            yh <- max(dsb$y)/20; yi <- yh/4
            lines(c(SB_LB,SB_LB,SB_LB,SB_UB,SB_UB,SB_UB),
                  c(yh-yi,yh+yi,yh   ,yh   ,yh-yi,yh+yi))
            text((SB_LB+SB_UB)/2, yh+yi*2, "Prediction band 95%")
          }
          pbin <- 0
          if (SB_LB<=0 & 0<=SB_UB) {pbin <- 1}
        }
      }
    }
  }
  m <- matrix(data=c(SB,SB_LB,SB_UB,pbin),nrow=1,ncol=4)
  colnames(m) <- c("SBadj","SBadj_LB","SBadj_UB","pbin")
  return(m)
}

\end{Verbatim}
which can be called by:\mynobreakpar
\begin{Verbatim}
# Agreement of Bell and Kato-Katz examination methods
Data <- ("
BellxKK P   N
P       184 54
N        14 63
")
m <- as.matrix(read.table(textConnection(Data),
                          header=TRUE, row.names=1))
print(m)
res <- agr2x2_badjShankar(m,B=100000,showboot=TRUE,plot=TRUE)
print(res)
\end{Verbatim}

\subsection{Implementation of Gwet's $AC1$}\label{ap:AC1}
The function \verb"agr2x2_ac1Gwet" implements equation~\ref{eq:AC1}, receiving the same parameters and returning a matrix alike \verb"agr2x2_kCohen"~(see~``\nameref{fnk_k}").  

Gwet's~$AC1$ is an index of agreement between observations, thus the higher the absolute value the greater is the agreement~(${AC1>0}$) or disagreement~(${AC1<0}$) between raters. 

One advantage of Gwet's~$AC1$ is her proposal of an estimator ballasted on a statistical test with a defined sample distribution, which leads to the computation of a $p$~value. This inferential statistics is implemented in R by \verb"irrCAC::gwet.ac1.raw", which is called from our implementation when \verb"test=TRUE" (this function requires a numeric data frame to process, using \verb"DescTools::Untable" and some additional transformations).

It was implemented by:\mynobreakpar
\begin{Verbatim}
# Gwet's AC1
agr2x2_ac1Gwet <- function(a,b,c,d,test=FALSE)
{
  if (is.matrix(a))
  {
    m <- a
    a <- m[1,1]
    b <- m[1,2]
    c <- m[2,1]
    d <- m[2,2]
  }
  ac1 <- (a^2 + d^2 - ((b+c)^2)/2 ) / 
    (a^2 + d^2 + ((b+c)^2)/2 + (a+d)*(b+c) )
  p.value <- NA
  if(is.finite(ac1))
  {
    if(test)
    {
      m <- matrix(data=c(a,b,c,d), nrow=2, ncol=2, byrow = TRUE)
      dt <- DescTools::Untable(m)
      dt$Var1 <- as.character(dt$Var1)
      dt$Var1[dt$Var1=="A"] <- "1"
      dt$Var1[dt$Var1=="B"] <- "0"
      dt$Var1 <- as.numeric(dt$Var1)
      dt$Var2 <- as.character(dt$Var2)
      dt$Var2[dt$Var2=="A"] <- "1"
      dt$Var2[dt$Var2=="B"] <- "0"
      dt$Var2 <- as.numeric(dt$Var2)
      ac1.test <- irrCAC::gwet.ac1.raw(dt)
      p.value <- ac1.test$est$p.value 
      if(!is.finite(p.value)) {p.value<-NA}
    }
  }
  m <- matrix(data=c(ac1,p.value),nrow=1,ncol=2)
  colnames(m) <- c("AC1","p")
  return(m)
}

\end{Verbatim}
For example:\mynobreakpar
\begin{Verbatim}
# Agreement of Bell and Kato-Katz examination methods
Data <- ("
BellxKK P   N
P       184 54
N        14 63
")
m <- as.matrix(read.table(textConnection(Data),
                          header=TRUE, row.names=1))
print(m)
res <- agr2x2_ac1Gwet(m,test=TRUE)
print(res)
\end{Verbatim}

\subsection{Comment on null hypotheses}\label{ap:H0}

McNemar's~$\chi^2$ testing is the only estimator whose null hypothesis is reverse, concluding for absence of change (i.e., ${b=c}$) when the $p$~value is significant (all other proposed estimators conclude for agreement or disagreement by rejection of their respective null hypotheses). 

For the Bell and Kato-Katz example applied by Kirkwood and Sterne, 2003, pp.~216-218~\cite{Kirkwood2003}, McNemar's~$\chi^2$ leads to rejection of the null hypothesis and the authors concluded for the disagreement between methods, stating superiority of @@. This is diverse from the other concurrent estimators applied in the current work suggesting that both methods are equivalent (without any judgement of superiority). The problem is that the misuse of McNemar's~$\chi^2$ that cannot be applied for agreement/disagreement decisions, as discussed in the main text.

It order to further emphasize that McNemar's~$\chi^2$ is not an agreement estimator we apply all implemented functions to a hypothetical matrix that show obvious agreement between two raters scoring a hypothetical measurement method as positive or negative, implemented by:\mynobreakpar
\begin{Verbatim}
# agr2x2_4tests.R

# required functions
source("agr2x2_kCohen.R")
source("agr2x2_qYule.R")
source("agr2x2_rPearson.R")
source("agr2x2_piScott.R")
source("agr2x2_f1Dice.R")
source("agr2x2_ac1Gwet.R")
source("agr2x2_SAC.R")
source("agr2x2_G.R")
source("agr2x2_chi2mnMcNemar.R")
source("agr2x2_mnMcNemar.R")
source("agr2x2_mnMcNemar2010.R")
source("agr2x2_mnMcNemar2017.R")

# hypothetical matrix
hyp_m <- matrix(data=c(70,2,4,40), nrow=2, ncol=2, byrow=TRUE)
rownames(hyp_m) <- c("Positive","Negative")
colnames(hyp_m) <- c("Positive","Negative")
print(hyp_m)

cat("\n---- Cohen's kappa ----\n")
prmatrix(agr2x2_kCohen(hyp_m,test=TRUE),rowlab="",quote=FALSE)
cat("\n---- Yule's Q ----\n")
prmatrix(agr2x2_qYule(hyp_m,test=TRUE),rowlab="",quote=FALSE)
cat("\n---- Pearson's r ----\n")
prmatrix(agr2x2_rPearson(hyp_m,test=TRUE),rowlab="",quote=FALSE)
cat("\n---- Scott's pi ----\n")
prmatrix(agr2x2_piScott(hyp_m,test=TRUE),rowlab="",quote=FALSE)
cat("\n---- Dice's F1 ----\n")
prmatrix(agr2x2_f1Dice(hyp_m,B=10000),rowlab="",quote=FALSE)
cat("\n---- Gwet's AC1 ----\n")
prmatrix(agr2x2_ac1Gwet(hyp_m,test=TRUE),rowlab="",quote=FALSE)
cat("\n---- SAC ----\n")
prmatrix(agr2x2_SAC(hyp_m,B=10000),rowlab="",quote=FALSE)
cat("\n---- Holley and Guilford's GC ----\n")
prmatrix(agr2x2_G(hyp_m,test=TRUE),rowlab="",quote=FALSE)
cat("\n---- Normalized McNemar's chi-squared ----\n")
prmatrix(agr2x2_mnMcNemar(hyp_m,B=10000),rowlab="",quote=FALSE)
cat("\n---- Traditional McNemar's chi-squared ----\n")
prmatrix(agr2x2_chi2mnMcNemar(hyp_m,test=TRUE),rowlab="",quote=FALSE)
cat("\n---- Rev. Lu (2010) McNemar's chi-squared ----\n")
prmatrix(agr2x2_mnMcNemar2010(hyp_m,B=10000),rowlab="",quote=FALSE)
cat("\n---- Rev. Lu et al. (2017) McNemar's chi-squared ----\n")
prmatrix(agr2x2_mnMcNemar2017(hyp_m,B=10000),rowlab="",quote=FALSE)
\end{Verbatim}

This procedure results in:\mynobreakpar
\begin{Verbatim}
         Positive Negative
Positive       70        2
Negative        4       40

---- Cohen's kappa ----
        k        kM            p
 0.889172 0.9630573 9.406842e-22

---- Yule's Q ----
        Q      p.value
 0.994302 4.765713e-24

---- Pearson's r ----
         r      p.value
 0.8897794 1.220942e-40

---- Scott's pi ----
        pi      piLB     piUB p_bin
 0.8891367 0.8014564 0.976817     0

---- Dice's F1 ----
        F1    F1.adj  F1.adjLB  F1.adjUB pbin
 0.9589041 0.9178082 0.9238724 0.9899084    0

---- Gwet's AC1 ----
       AC1 p
 0.9030371 0

---- SAC ----
       SAC     SACLB     SACUB pbin
 0.8965517 0.8081249 0.8190539    0

---- Holley and Guilford's GC ----
         G p
 0.8965517 0

---- Normalized McNemar's chi-squared ----
        MN       MN.LB     MN.UB pbin
 0.3333333 -0.05950388 0.9329642    1

---- Traditional McNemar's chi-squared ----
    Chi2MN b/c     b/c.LB   b/c.UB      p
 0.6666667 0.5 0.04522901 3.488772 0.6875

---- Rev. Lu (2010) McNemar's chi-squared ----
    MN2010   MN2010LB MN2010UB pbin2010
 0.4113475 -0.3105407 3.280514        1

---- Rev. Lu et al. (2017) McNemar's chi-squared ----
     MN2017    MN2017LB  MN2017UB pbin2017
 0.03695444 -0.03148964 0.3442073        1
\end{Verbatim}

Observe the similarity of the point-estimate of all estimators and the rejection of null hypothesis. All variants of McNemar's~$\chi^2$ are discrepant regarding both estimation value and $p$~value. A McNemar's test does not measure agreement, but tests ${H_0:b=c}$, which is applicable to pre-post situations. For instance, after and before an intervention (for instance, a electoral debate) it can be applied to verify changes of opinions from `-' to `+'~($b$) or `+' to `-'~($c$) of a group of voters.

\section{Computation of Tables~\ref{tab:challenge} and \ref{tab:extremes}}\label{ap:challenge}

In order to create challenge scenarios, two worksheets (Excel format) with 2x2~configurations are previously stored in folder \verb"data". 
After processing, the results are stored in other two worksheets in folder \verb"result", from which data were transcribed to Tables~\ref{tab:challenge} and \ref{tab:extremes}.
Respectively:
\begin{itemize}
\item The choice of tables are in folder \verb"data": 
	\subitem \verb"agr2x2_table3_input.xlsx" and
	\subitem \verb"agr2x2_table4_input.xlsx".
\item Results are stored in folder \verb"result": 
		\subitem \verb"agr2x2_table3_output.xlsx" and 
		\subitem \verb"agr2x2_table4_output.xlsx".
\end{itemize}

Processing was implemented with \verb"agr2x2_tables_3_4.R":\mynobreakpar
\begin{Verbatim}
# Dependencies (libraries):
fileinput <- file.path("data",c("agr2x2_table3_input.xlsx",
                                "agr2x2_table3_input.xlsx"))
fileoutput <- file.path("result",c("agr2x2_table3_output.xlsx",
                                   "agr2x2_table4_output.xlsx"))
dir.create("result", showWarnings=FALSE)

for (t.aux in 1:length(fileinput))
{
  data <- readxl::read_excel(fileinput[t.aux])
  # Holley and Guilford's G
  source("agr2x2_G.R")
  data$G <- NA
  # Simple Agreement Coefficient
  # (same as G with CB95% by bootstrapping)
  source("agr2x2_SAC.R")
  data$SAC <- NA
  # Gwet's AC1
  source("agr2x2_ac1Gwet.R")
  data$AC1 <- NA
  # Scott's pi
  source("agr2x2_piScott.R")
  data$pi <- NA
  # Cohen's kappa
  source("agr2x2_kCohen.R")
  data$k <- NA
  data$kM <- NA
  # Pearson's r
  source("agr2x2_rPearson.R")
  data$r <- NA
  # Yule's Q
  source("agr2x2_qYule.R")
  data$Q <- NA
  # Yule's Y
  source("agr2x2_yYule.R")
  data$Y <- NA
  # Hubert's GH
  source("agr2x2_gHubert.R")
  data$GH <- NA
  # Bangdiwala and Shankar's SB
  source("agr2x2_bShankar.R")
  data$B <- NA
  source("agr2x2_badjShankar.R")
  data$Badj <- NA
  # Dice's F1
  source("agr2x2_f1Dice.R")
  data$F1 <- NA
  data$F1adj <- NA
  # McNemar's chi-squared
  source("agr2x2_mnMcNemar.R")
  data$MN <- NA
  # Traditional McNemar's chi-squared
  source("agr2x2_chi2mnMcNemar.R")
  data$Chi2MN <- NA
  # McNemar's chi-squared revised by Yu (2010, 2017))
  source("agr2x2_mnMcNemar2010.R")
  data$MN2010 <- NA
  source("agr2x2_mnMcNemar2017.R")
  data$MN2017 <- NA
  # other files required (bootstrapping auxiliar)
  source("agr2x2_boot.table.R")

  cat("\nComputing ",fileinput[t.aux],"\n",sep="")
  for (r.aux in 1:nrow(data))
  {
    a <- data$a[r.aux]
    b <- data$b[r.aux]
    c <- data$c[r.aux]
    d <- data$d[r.aux]
    m <- matrix(data=c(a,b,c,d), nrow=2, ncol=2, byrow = TRUE)
    # Holley and Guilford's G
    res <- agr2x2_G(m)
    data$G[r.aux] <- as.numeric(res[1])
    # Simple Agreement Coefficient
    # (same as G with CB95% by bootstrapping)
    res <- agr2x2_SAC(m)
    data$SAC[r.aux] <- as.numeric(res[1])
    # Gwet's AC1
    res <- agr2x2_ac1Gwet(m)
    data$AC1[r.aux] <- as.numeric(res[1])
    # Scott's pi
    res <- agr2x2_piScott(m)
    data$pi[r.aux] <- as.numeric(res[1])
    # Cohen's kappa
    res <- agr2x2_kCohen(m)
    data$k[r.aux] <- as.numeric(res[1])
    data$kM[r.aux] <- as.numeric(res[2])
    # Pearson's r
    res <- agr2x2_rPearson(m)
    data$r[r.aux] <- as.numeric(res[1])
    # Yule's Q
    res <- agr2x2_qYule(m)
    data$Q[r.aux] <- as.numeric(res[1])
    # Yule's Y
    res <- agr2x2_yYule(m)
    data$Y[r.aux] <- as.numeric(res[1])
    # Hubert's Gamma
    res <- agr2x2_gHubert(m)
    data$GH[r.aux] <- as.numeric(res[1])
    # Bangdiwala and Shankar's B
    res <- agr2x2_bShankar(m)
    data$B[r.aux] <- as.numeric(res[1])
    res <- agr2x2_badjShankar(m)
    data$Badj[r.aux] <- as.numeric(res[1])
    # Dice's F1
    res <- agr2x2_f1Dice(m)
    data$F1[r.aux] <- as.numeric(res[1])
    data$F1adj[r.aux] <- as.numeric(res[2])
    # Normalized McNemar's chi-squared
    res <- agr2x2_mnMcNemar(m)
    data$MN[r.aux] <- as.numeric(res[1])
    # McNemar's chi-squared revised by Yu (2010, 2017))
    res <- agr2x2_chi2mnMcNemar(m)
    data$Chi2MN[r.aux] <- as.numeric(res[1])
    res <- agr2x2_mnMcNemar2010(m)
    data$MN2010[r.aux] <- as.numeric(res[1])
    res <- agr2x2_mnMcNemar2017(m)
    data$MN2017[r.aux] <- as.numeric(res[1])
  }
  openxlsx::write.xlsx(data,fileoutput[t.aux])
  cat("\rResults stored in ",fileoutput[t.aux],"\n",sep="")
}
\end{Verbatim}

McNemar's~$\chi^2$~\cite{McNemar1947} and revisions~\cite{Lu2010,Lu2017} were not included in these two tables for the reasons explained in the main text.

\section{Computation of Table~\ref{tab:correlation}}\label{ap:r_hdi}

In order to obtain correlations between estimatives, the file \verb"from1to68.csv" stored in folder \verb"result" is required, already containing all estimatives computed (see~``\nameref{ap:fig_hexbin}").

The procedure \verb"agr2x2_correlations.R" computes Pearson's and Spearman's correlations of Holley and Guilford's~$G$ against all other estimators, creating three files in folder \verb"result": 
\begin{itemize}
\item \verb"agr2x2_correlations.csv" --- all computed correlations.
\item \verb"agr2x2_tablehdi.csv" --- data summary, transcribed to Table~\ref{tab:correlation}.
\item \verb"agr2x2_correlations.pdf" --- PDF file containing:
	\begin{itemize}
	\item graphs of correlation values ($r$ and $\rho$) with confidence interval 95\% estimated by the R function \verb"cor.test" in function of $1 \le n \le 68$.
	\item multiple scatterplots (analogous to Figure~\ref{fig:hexbin1to68}) to verify the stability of each estimator with $1 \le n \le 68$. To these scatterplots it was added the bisector (dashed line) and a robust trend line (solid line from function \verb"lowess"); the greater the relative countings the increasing size and redish color of the bullets.
	\end{itemize}
\end{itemize}

The implementation follows:\mynobreakpar
\verb"agr2x2_correlations.R": \mynobreakpar
\begin{Verbatim}
# agr2x2_correlations.R
folder_out <- "result"
dir.create(folder_out, showWarnings=FALSE)

source("eiras.redblue.gradation.R")

from <- 1
to <- 68
filename <- file.path("result",paste0("from",from,"to",to,".csv"))
data <- as.data.frame(data.table::fread(filename,header=TRUE))
cat("\nRead ",filename,"\n",sep="")

# columns of interest
colbase <- "G"
extcolbase <- "Holley and Guilford's G"
colnames <- c("AC1",
              "pi",
              "k", "kM",
              "r", "Q", "Y",
              "SB", "SBadj",
              "F1", "F1adj",
              "MN",
              "Chi2MN", "MN2010", "MN2017")
extcolnames <- c("Gwet's AC1",
                 "Scott's pi",
                 "Cohen's kappa",
                 "Corrected kappa",
                 "Pearson's r", "Yule's Q", "Yule's Y",
                 "Bangdiwala and Shankar's B", "Adjusted B",
                 "Dice's F1", "Adjusted F1",
                 "Norm. McNemar's Chi-squared",
                 "McNemar's Chi-squared",
                 "Revised 2010 McNemar's Chi-squared",
                 "Revised 2017 McNemar's Chi-squared"
)

dt_res <- data.frame(from:to)
names(dt_res) <- "n"
for (c.aux in 1:length(colnames))
{
  dt_res$new1 <- NA
  dt_res$new2 <- NA
  dt_res$new3 <- NA
  dt_res$new4 <- NA
  dt_res$new5 <- NA
  dt_res$new6 <- NA
  dt_res$new7 <- NA
  dt_res$new8 <- NA
  dt_res$new9 <- NA
  names(dt_res) <- c(names(dt_res)[1:(ncol(dt_res)-9)],
                     colnames[c.aux],
                     paste0("r_",colnames[c.aux]),
                     paste0("r.p_",colnames[c.aux]),
                     paste0("rLB_",colnames[c.aux]),
                     paste0("rUBL_",colnames[c.aux]),
                     paste0("rho_",colnames[c.aux]),
                     paste0("rho.p_",colnames[c.aux]),
                     paste0("rhoLB_",colnames[c.aux]),
                     paste0("rhoUB_",colnames[c.aux])
  )
}

cat("\nComputing correlations...\n")
rotate <- c("/","-","\\","|")
idxrotate <- 1
for (n in from:to)
{
  c.num <- which(names(data)==colbase)
  for (c.aux in 1:length(colnames))
  {
    cat("\r",rotate[idxrotate],sep="")
    idxrotate <- idxrotate+1; if(idxrotate>length(rotate)){idxrotate<-1}
    c2.dat <- which(names(data)==colnames[c.aux])
    c2.res <- which(names(dt_res)==colnames[c.aux])
    x <- as.numeric(unlist(data[data$n==n,c.num]))
    y <- as.numeric(unlist(data[data$n==n,c2.dat]))
    res <- NA
    try (
      res <- cor.test(x,y),
      silent=TRUE
    )
    if(!is.na(res[1]))
    {
      dt_res[dt_res$n==n,c2.res  ] <- 1 # it is only to locate columns
      dt_res[dt_res$n==n,c2.res+1] <- res$estimate
      if (!is.null(res$conf.int))
      {
        dt_res[dt_res$n==n,c2.res+2] <- res$p.value
        dt_res[dt_res$n==n,c2.res+3] <- res$conf.int[1]
        dt_res[dt_res$n==n,c2.res+4] <- res$conf.int[2]
      }
    }
    x <- rank(x)
    y <- rank(y)
    res <- NA
    try (
      res <- cor.test(x,y),
      silent=TRUE
    )
    if(!is.na(res[1]))
    {
      dt_res[dt_res$n==n,c2.res+5] <- res$estimate
      if (!is.null(res$conf.int))
      {
        dt_res[dt_res$n==n,c2.res+6] <- res$p.value
        dt_res[dt_res$n==n,c2.res+7] <- res$conf.int[1]
        dt_res[dt_res$n==n,c2.res+8] <- res$conf.int[2]
      }
    }
  }
}
corfile <- file.path(folder_out,"agr2x2_correlations.csv")
data.table::fwrite(dt_res,corfile)

# median and high density intervals of estimators
nomeshdi <- c("Estimator","r.Median","r.HDI_LB","r.HDI_UB",
              "rho.Median","rho.HDI_LB","rho.HDI_UB")
dt_hdi <- data.frame(matrix(nrow=length(colnames),ncol=length(nomeshdi)))
names(dt_hdi) <- nomeshdi
dt_hdi$Estimator <- colnames
for (r.aux in 1:nrow(dt_hdi))
{
  # Pearson's r
  c.num <- which(names(dt_res)==paste0("r_",dt_hdi$Estimator[r.aux]))
  v <- as.numeric(unlist(dt_res[,c.num]))
  dt_hdi$r.Median[r.aux] <- median(v, na.rm=TRUE)
  d <- density(v, na.rm=TRUE)
  hdi <- HDInterval::hdi(d, credMass=0.95, allowSplit = FALSE)
  dt_hdi$r.HDI_LB[r.aux] <- as.numeric(hdi[1])
  dt_hdi$r.HDI_UB[r.aux] <- as.numeric(hdi[2])
  if (hdi[1] < -1){hdi[1] <- -1}
  if (hdi[2] >  1){hdi[2] <-  1}
  # Spearman's rho
  c.num <- which(names(dt_res)==paste0("rho_",dt_hdi$Estimator[r.aux]))
  v <- as.numeric(unlist(dt_res[,c.num]))
  dt_hdi$rho.Median[r.aux] <- median(v, na.rm=TRUE)
  d <- density(v, na.rm=TRUE)
  hdi <- HDInterval::hdi(d, credMass=0.95, allowSplit = FALSE)
  if (hdi[1] < -1){hdi[1] <- -1}
  if (hdi[2] >  1){hdi[2] <-  1}
  dt_hdi$rho.HDI_LB[r.aux] <- as.numeric(hdi[1])
  dt_hdi$rho.HDI_UB[r.aux] <- as.numeric(hdi[2])
}
for (c.aux in 2:ncol(dt_hdi))
{
  dt_hdi[,c.aux] <- round(dt_hdi[,c.aux],4)
}
# order by median Spearman's coefficient
dt_hdi <- dt_hdi[order(dt_hdi$rho.Median, decreasing=TRUE),]
print(dt_hdi)
hdifile <- file.path(folder_out,"agr2x2_tablehdi.csv")
data.table::fwrite(dt_hdi,hdifile)

# graphs
pdffile <- file.path(folder_out,"agr2x2_correlations.pdf")
pdf(pdffile,paper="letter")
cat("\nProcessing graphs...\n")
o.par <- par()
m <- matrix(c(1,1,2:69),ncol=10,nrow=7,byrow=TRUE)
# assuming best estimator as reference
c.num <- which(names(data)==colbase)
for (i.aux in 1:nrow(dt_hdi))
{
  c.aux <- which(colnames==dt_hdi$Estimator[i.aux])
  cat("\t",extcolnames[c.aux],"\n")

  # correlations in function of n
  c2.dat <- which(names(data)==colnames[c.aux])
  plot (dt_res$n, as.numeric(unlist(dt_res[,c2.res+1])),
        main=extcolnames[c.aux],
        xlab="n", ylab="r", ylim=c(-1,1),
        pch=21, col="black", bg="black")
  for (r.aux in 1:nrow(dt_res))
  {
    lines(rep(dt_res$n[r.aux],2),
          c(as.numeric(unlist(dt_res[r.aux,c2.res+3])),
            as.numeric(unlist(dt_res[r.aux,c2.res+4]))))
  }
  plot (dt_res$n, as.numeric(unlist(dt_res[,c2.res+5])),
        main=extcolnames[c.aux],
        xlab="n", ylab="rho", ylim=c(-1,1),
        pch=21, col="black", bg="black")
  for (r.aux in 1:nrow(dt_res))
  {
    lines(rep(dt_res$n[r.aux],2),
          c(as.numeric(unlist(dt_res[r.aux,c2.res+7])),
            as.numeric(unlist(dt_res[r.aux,c2.res+8]))))
  }

  # map of estimators by each size n
  c2.dat <- which(names(data)==colnames[c.aux])
  ymin <- min(as.numeric(unlist(data[,c2.dat])),na.rm=TRUE)
  range.ymin <- ymin
  if(ymin > -1){ymin <- -1} # for graph scale
  ymax <- max(as.numeric(unlist(data[,c2.dat])),na.rm=TRUE)
  range.ymax <- ymax
  if(ymax <  1){ymax <-  1} # for graph scale
  layout(m)
  par(mai=c(0,0,0,0))
  plot(NA,axes=FALSE,xlim=c(0,12), ylim=c(0,6),xlab="",ylab="")
  text(6,5,paste0(extcolnames[c.aux],"\n(y axis)\n",
                  "vs.\n",
                  colbase,"\n(x axis)\n\n",
                  "range(G): [-1,1]\n",
                  "range(",colnames[c.aux],"): [",
                  range.ymin,",",range.ymax,"]"),
       cex=0.85, pos=1)
  par(mai=c(0.2,0.2,0.2,0))
  for (n in from:to)
  {
    x <- round(as.numeric(unlist(data[data$n==n,c.num])),2)
    y <- round(as.numeric(unlist(data[data$n==n,c2.dat])),2)
    dt_xy <- as.data.frame(table(x,y))
    dt_xy <- dt_xy[dt_xy$Freq>0,]
    dt_xy$x <- as.numeric(as.character(dt_xy$x))
    dt_xy$y <- as.numeric(as.character(dt_xy$y))
    plot(NA,
         main=paste0("n=",n), xlab="", ylab="",
         xlim=c(-1.1,1.1),
         ylim=c(ymin-abs((ymax-ymin)/10),
                ymax+abs((ymax-ymin)/10)),
         axes=FALSE)
    if(nrow(dt_xy)>0)
    {
      max <- max(dt_xy$Freq,na.rm=TRUE)
      dt_xy$FreqR <- dt_xy$Freq/max
      dt_xy$cexR <- 0.2+2*dt_xy$FreqR
      dt_xy <- dt_xy[order(dt_xy$FreqR),]
      for(i.x in 1:nrow(dt_xy))
      {
        points(dt_xy$x[i.x],dt_xy$y[i.x],
               pch=21, cex=dt_xy$cexR[i.x],
               col="transparent",
               bg=paste0(redblue.gradation(dt_xy$FreqR[i.x]),"44"))
      }
      # bisector (if applicable)
      if(ymax==1)
      {
        lines(c(-1,1),c(-1,1),lwd=2.0,col="white")
        lines(c(-1,1),c(-1,1),lwd=0.5,lty=2,col="black")
      }
      # robust trend line
      ll <- lowess(dt_xy$x,dt_xy$y)
      lines(ll,lwd=3.6,col="white")
      lines(ll,lwd=1.0,col="black")
    }
  } # for n
  par(o.par)
} # for i.aux
dev.off()
cat("\n")
cat("Median and HDI95% stored in ",hdifile,"\n",sep="")
cat("All raw correlations stored in ",corfile,"\n",sep="")
cat("Graphs stored in ",pdffile,"\n",sep="")
\end{Verbatim}

The script \verb"agr2x2_correlations.R" depends on other three auxiliary scripts for the graph colors that my be easily replaced according to the user's taste:\mynobreakpar
\begin{itemize}
\item \verb"eiras.redblue.gradation.R"
\begin{Verbatim}
# eiras.redblue.gradation.R
# receives a hot[0,1] level and returns rgb

source("eiras.rgb2rgbstring.R")

redblue.gradation <- function (hot, color=TRUE)
{
  # hot ... r
  #  1  ... 255
  if (hot<0) {hot <- 0}
  if (hot>1) {hot <- 1}
  if (color==TRUE)
  {
    r <- 255*hot
    if (r<(255/2))
    {
      g <- r
    } else
    {
      g <- 150-r/2.5
    }
    b <- 170-r/1.5 # (256/4)-r/4
  } else
  {
    r <- 60+(210-80)*(1-hot)
    g <- b <- r
  }
  return (rgb2rgbstring(r,g,b))
}
\end{Verbatim}
\item \verb"eiras.rgb2rgbstring.R"
\begin{Verbatim}
# eiras.rgb2rgbstring.R

library(grDevices)
source("eiras.text.leading.R")

rgb2rgbstring <- function(r, g, b, a)
{
  r <- round(r,0)
  g <- round(g,0)
  b <- round(b,0)
  a <- try(round(a,0), silent = TRUE)
  r <- text.leading(as.character(as.hexmode(as.numeric(r))),2,"0")
  g <- text.leading(as.character(as.hexmode(as.numeric(g))),2,"0")
  b <- text.leading(as.character(as.hexmode(as.numeric(b))),2,"0")
  if (is.numeric(a))
  {
    a <- text.leading(as.character(as.hexmode(as.numeric(a))),2,"0")
  } else
  {
    a <- ""
  }
  return(paste("#",r,g,b,a,sep=""))
}
\end{Verbatim}
\item \verb"eiras.text.leading.R"
\begin{Verbatim}
# eiras.text.leading.R
# returns text with leading characters

text.leading <- function (text, lentxt, lead.chr=" ")
{
  text <- sprintf("%s",text)
  while(nchar(text) < lentxt)
  {
    text <- paste(lead.chr,text,sep="")
  }
  return(text)
}
\end{Verbatim}
\end{itemize}

\section{Creating all 2x2 tables with size $n$}\label{ap:n1to68}

It was implemented \verb"agr2x2_gentablen", a function to create all possible 2x2~tables of size~\verb"n":\mynobreakpar
\begin{Verbatim}
# create a list of tables of size n
agr2x2_gentablen <- function(n)
{
  t <- 0
  list_tables <- list( )
  for(i in 0:n)
  {
    for(j in 0:(n-i)) 
    {
      for (k in 0:(n-i-j)) 
      {
        t <- t + 1
        vetor <- c(i,j,k,n-i-j-k)
        list_tables[[t]] <- vetor
      }
    }
  }
  return(list_tables)
}
\end{Verbatim}

This function returns a list of tables and it is easy to use throught a coordinator function that can apply a range of \verb"n" values (\verb"from"...\verb"to") concatenated in a more convenient data frame and stored in disk as a \verb"csv" file:\mynobreakpar
\begin{Verbatim}
# agr2x2_createtables.R

source("agr2x2_gentablen.R")

agr2x2_createtables <- function(from, to)
{
  # creating all tables from ... to
  all_n <- from:to
  table_acm <- c("a","b","c","d")
  dt_tables_acm <- data.frame(matrix(nrow=0,ncol=4))
  names(dt_tables_acm) <- table_acm
  for (n in all_n)
  {
    tables<-agr2x2_gentablen(n)
    dt_tables <- data.frame(matrix(data=unlist(tables),
                                   ncol=4,byrow=TRUE))
    names(dt_tables) <- table_acm
    cat("\nn = ",n," ... ",nrow(dt_tables)," tabelas validas",sep="")
    dt_tables_acm <- rbind(dt_tables_acm,dt_tables)
  }
  dt_tables_acm$n <- dt_tables_acm$a+dt_tables_acm$b+
                     dt_tables_acm$c+dt_tables_acm$d
  # results in csv
  filename <- file.path("data",paste0("from",from,"to",to,".csv"))
  data.table::fwrite(dt_tables_acm,filename)
  cat("\nResults stored in ",filename,"\n")
  return(filename)
}
\end{Verbatim}

For instance, to obtain all 2x2~tables with $n=2$ and $n=3$:\mynobreakpar
\begin{Verbatim}
source("agr2x2_createtables.R")

tables <- as.data.frame(
  data.table::fread(agr2x2_createtables(from=2,to=3),header=TRUE)
                       )
\end{Verbatim}

In this example, \verb"agr2x2_createtables" creates the filename \verb"from2to3.csv", which contains all required tables for future applications; the filename can be recovered with \verb"data.table::fread" and stored in a data frame, \verb"tables", containing:\mynobreakpar
\begin{Verbatim}
   a b c d n
1  0 0 0 2 2
2  0 0 1 1 2
3  0 0 2 0 2
4  0 1 0 1 2
5  0 1 1 0 2
6  0 2 0 0 2
7  1 0 0 1 2
8  1 0 1 0 2
9  1 1 0 0 2
10 2 0 0 0 2
11 0 0 0 3 3
12 0 0 1 2 3
13 0 0 2 1 3
14 0 0 3 0 3
15 0 1 0 2 3
16 0 1 1 1 3
17 0 1 2 0 3
18 0 2 0 1 3
19 0 2 1 0 3
20 0 3 0 0 3
21 1 0 0 2 3
22 1 0 1 1 3
23 1 0 2 0 3
24 1 1 0 1 3
25 1 1 1 0 3
26 1 2 0 0 3
27 2 0 0 1 3
28 2 0 1 0 3
29 2 1 0 0 3
30 3 0 0 0 3
\end{Verbatim}

\section{Computation of Figure~\ref{fig:densityplots64}}\label{ap:n64}

\subsection{Creation of all 2x2 tables with $n$=64}\label{ap:n64_2x2}
We created a file containing all 47,905 possible tables with 

\verb"agr2x2_createtables(from=64,to=64)"

whose results are stored in \verb"from64to64.csv" (folder \verb"data", see~``\nameref{ap:n1to68}" for 2x2~table creation).

\subsection{Inferential tests}\label{ap:inferential}
Using the functions described in this supplemental material (see~``\nameref{ap:Rfunctions}"), the following R script computed and stored all inferential results using \verb"agr2x2_main.R":\mynobreakpar
\VerbatimInput[frame=single, fontsize=\footnotesize]{agr2x2_main.R}

To the resulting file, \verb"from64to64.csv" (folder~\verb"result"), columns were added to store the computation of all estimators described in the supplemental material,~``\nameref{ap:Rfunctions}".

\subsection{Figures and performance checking}\label{ap:mistakes}
To generate figures, \verb"from64to64.csv" (stored in folder~\verb"result") must contain all results of the inferential tests processed in the previous step~(``\nameref{ap:inferential}"). Taking each studied estimator as if it was a benchmark, it generates a table with the interval of $\frac{a+d}{n}$ that corresponds to the non-rejection of the null hypothesis (the lower and upper bounds, the range and the center of this intervals are stored), it computes the mean number of \verb"Total" mistakes and the \verb"Global" mean mistakes. The adopted benchmark must minimize the \verb"Global" mistakes and should have the narrowest interval centered in $\frac{a+d}{n} = 0.5$. 

This function was implemented in\newline
\verb"agr2x2_densitygraphs.R":\mynobreakpar
\VerbatimInput[frame=single, fontsize=\footnotesize]{agr2x2_densitygraphs.R}

This function depends, for the construction of individual graphs, of\newline
\verb"agr2x2_diffgraph.R":\mynobreakpar
\VerbatimInput[frame=single, fontsize=\footnotesize]{agr2x2_diffgraph.R}

In order to test all estimators as possible benchmarks, it is called by\mynobreakpar
\begin{Verbatim}
# agr2x2_densitygraphs_example.R
# trying each estimator as benchmark to all others

source("agr2x2_densitygraphs.R")

n <- 64
filename <- file.path("result",paste0("from",n,"to",n,".csv"))

# trying any estimator as benchmark
coldata <- c("G",
             "SAC",
             "AC1",
             "pi",
             "k",
             "r",
             "Y", "Q",
             "SBadj", "SB",
             "F1",
             "MN",
             "Chi2MN",
             "MN2010",
             "MN2017"
)
coltitles <- c("Holley and Guilford's G",
               "Simple Agreement Coefficient",
               "Gwet's AC1" ,
               "Scott's pi",
               "Cohen's kappa",
               "Pearson's r",
               "Yule's Y", "Yule's Q",
               "Bangdiwala and Shankar's B (rescaled)",
               "Bangdiwala and Shankar's B",
               "Dice's F1",
               "Normalized McNemar's Chi-squared",
               "Traditional McNemar's Chi-squared",
               "McNemar's Chi-squared rev. 2010",
               "McNemar's Chi-squared rev. 2017"
)

# Show benchmark table
filemistakes <- agr2x2_densitygraphs(filename,
                                     colbase=coldata,
                                     colnames=coldata,
                                     titles=coltitles, format="png")
benchmark <- as.data.frame(data.table::fread(filename,header=TRUE))
prmatrix(round(benchmark,4),rowlab=rep("",nrow(benchmark)),quote=FALSE)
\end{Verbatim}

This script stores figures in \verb"png" format (folder \verb"image"), taking each test as benchmark to all others. All combinations provide countings of mistakes, which are stored in \verb"agr2x2_mistakes_n64.csv" (folder~\verb"result"). From this procedures $G$ was the one that minimized the mistakes of all others and, for that reason, Holley and Guilford's~$G$ is the choice as benchmark and the image

\verb"agr2x2_densityplots_n64_G.png"

was elected as Figure~\ref{fig:densityplots64}~(the traditional McNemar's~$\chi^2$ and its two revisions were suppressed for the reasons explained in the main text).

\section{Computation of Figures~\ref{fig:hexbin1to68},~\ref{fig:hexbin_AC1},~\ref{fig:hexbin_k},~and~\ref{fig:hexbin_MN}}\label{ap:fig_hexbin}

The generation of Figures~\ref{fig:hexbin1to68},~\ref{fig:hexbin_AC1},~\ref{fig:hexbin_k},~and~\ref{fig:hexbin_MN} requires all estimators previously computed for all possible tables with $1 \le n \le 68$ (a total of 1,028,789~tables), using

\verb"agr2x2_createtables(from=1,to=68)" (see~``\nameref{ap:n1to68}")

which must store the resulting file in the folder \verb"data". Then, we applied the same R script described in this supplemental material, ~section~``\nameref{ap:n64}", subsection~\nameref{ap:inferential}", changing the initial parameters (the inferential tests are not required) to:
\begin{verbatim}
from <- 1
to <- 68
bootstraps <- FALSE
test <- FALSE
\end{verbatim}

which add the necessary columns to store the computation of all estimators (described in~``\nameref{ap:Rfunctions}"). This processed file \verb"from1to68.csv" (now stored in folder \verb"result" is the basis for Figures~\ref{fig:hexbin1to68},~\ref{fig:hexbin_AC1},~\ref{fig:hexbin_k},~and~\ref{fig:hexbin_MN}.

These figures are generated with hexbins applying the packages \verb"ggplot2" and \verb"ggpubr", a more efficient alternative to plot huge amounts of data than regular scatterplots.

Figure~\ref{fig:hexbin1to68} was generated with\newline
\verb"agr2x2_scatterplot_multi.R":\mynobreakpar
\VerbatimInput[frame=single, fontsize=\footnotesize]{agr2x2_scatterplot_multi.R}
The resulting figure is stored in \verb"agr2x2_scatterplot_multi.pdf" (folder \verb"image").

The detailed Figures~\ref{fig:hexbin_AC1},~\ref{fig:hexbin_k},~and~\ref{fig:hexbin_MN} can also be generated in folder \verb"image" with\newline
\verb"agr2x2_scatterplot_cases.R"\mynobreakpar
\VerbatimInput[frame=single, fontsize=\footnotesize]{agr2x2_scatterplot_cases.R}

This script generates all estimator cases, from which we selected to the main text:\mynobreakpar
\begin{itemize}
\item Figure~\ref{fig:hexbin_AC1}: \verb"agr2x2_scatterplot_case_AC1.pdf"
\item Figure~\ref{fig:hexbin_k}: \verb"agr2x2_scatterplot_case_k.pdf"
\item Figure~and~\ref{fig:hexbin_MN}: \verb"agr2x2_scatterplot_case_MN.pdf"

\end{itemize}

\end{document}